\documentclass[12pt]{iopart}
\usepackage{epsfig}

\newcommand{\ben}{\begin{equation}}
\newcommand{\een}{\end{equation}}

\newcommand{\gtrsim}{\,\rlap{\lower3.7pt\hbox{$\mathchar\sim$}}
\raise1pt\hbox{$>$}\,}
\newcommand{\lesssim}{\,\rlap{\lower3.7pt\hbox{$\mathchar\sim$}}
\raise1pt\hbox{$<$}\,}

\newcommand{\be}{\begin{equation}}
\newcommand{\ee}{\end{equation}}  
\newcommand{\bea}{\begin{eqnarray}}
\newcommand{\eea}{\end{eqnarray}}  
\newcommand{\gag}{g_{a\gamma}}

\begin{document}

\title{Axion-like particle effects on the polarization of cosmic high-energy gamma sources}

\author{Nicola Bassan}
\address{SISSA \\ Via Beirut 2-4, 34014 Trieste, Italy\\}

\author{Alessandro Mirizzi}
\address{II. Institut f\"ur theoretische Physik, Universit\"at Hamburg,\\
Luruper Chaussee 149, 22761 Hamburg, Germany}        
 
\author{Marco Roncadelli}
\address{INFN, Sezione di Pavia\\ Via A.~Bassi 6, 27100 Pavia,
Italy\\}

\begin{abstract}
Various satellite-borne missions are being planned to measure the polarization of a large number of gamma-ray bursts (GRBs). We show 
that the polarization pattern resulting from the current models of GRB emission can be drastically modified by the existence of very light axion-like particles (ALPs), which are predicted by many extensions of the Standard Model of particle physics. Basically, the propagation of photons emitted by a GRB 
through cosmic magnetic fields with a domain-like structure induces photon-ALP mixing, which is expected to produce a strong modification of the initial photon polarization. Because of the random orientation of the magnetic field in each domain, this effect strongly depends  on the orientation of the line of sight. As a consequence, photon-ALP conversion considerably broadens the initial polarization distribution. Searching for such a  peculiar feature through future high-statistics polarimetric measurements therefore offers a new opportunity to discover very light ALPs.
\end{abstract}

\maketitle

%%%%%%%%%%%%%%%%%%%%%%%%%%%%%%%%%%%%%%%%%%%%%%%%%%%%%%%%%%%%%%%%%%%%%%
\section{Introduction} %%%%%%%%%%%%%%%%%%%%%%%%%%%%%%%%%%%%%%%%%%%%%%%
%%%%%%%%%%%%%%%%%%%%%%%%%%%%%%%%%%%%%%%%%%%%%%%%%%%%%%%%%%%%%%%%%%%%%%
A generic feature of many extensions of the Standard Model is the prediction of axion-like particles (ALPs), namely very light spin-zero bosons characterized by a two-photon coupling $a\gamma\gamma$. Besides occurring   in four-dimensional models~\cite{masso1,masso2,coriano1,coriano2}, ALPs naturally arise within compactified Kaluza-Klein theories~\cite{kk} as well as in superstring theories~\cite{Svrcek:2006yi,Arvanitaki:2009fg}. 
As the name itself suggests, ALPs are a  generalization of the axion, the pseudo-Goldstone boson associated with the Peccei-Quinn symmetry proposed as a natural solution to the strong 
CP problem~\cite{Peccei:1977hh, Peccei:1977ur,Weinberg:1977ma,Wilczek:1977pj,Kim:2008hd}. 
Specifically, while the axion is characterized by a strict relationship between its mass and  $a\gamma\gamma$ coupling constant, these two parameters are to be regarded 
as independent for ALPs~\cite{Masso:2006id}.
Depending on the actual values of their mass and $a\gamma\gamma$ coupling constant, ALPs can play an important
 role in cosmology, either as cold dark matter particles responsible for the structure formation in the Universe~\cite{cdm} or as quintessential dark energy~\cite{carroll} which presumably triggers the present accelerated cosmic expansion (see~\cite{Jaeckel:2010ni} for a recent review on ALPs).

Owing to the $a\gamma\gamma$ coupling, in an external electromagnetic field the phenomenon of photon-ALP mixing takes place\footnote{We stress that the external field is necessary in order to compensate for the photon-ALP spin mismatch.}. More specifically, two very interesting effects arise in such a situation. One is photon-ALP conversion (oscillation)~\cite{Raffelt:1987im,sikivie,Anselm:1987vj}, which  is exploited by the ADMX experiment to search for ALP dark matter~\cite{Duffy:2006aa}, by CAST to search for solar axions~\cite{Zioutas:2004hi,Andriamonje:2007ew,Arik:2008mq} and by the regeneration laser experiments~\cite{Robilliard:2007bq,Chou:2007zzc,Afanasev:2008jt,Fouche:2008jk,Ehret:2009sq,Ehret:2010mh,axion2010}. The other effect consists in the change of the polarization state of photons traveling in a magnetic field. In particular, an initially linearly polarized photon beam propagating in a transverse magnetic field acquires an elliptical polarization with the major axis rotated with respect to the direction of the initial 
polarization~\cite{Raffelt:1987im,Maiani:1986md,Gasperini:1987da}. A claim for a positive observational evidence of such an effect by the PVLAS collaboration~\cite{Zavattini:2005tm} employing a laser beam has subsequently been withdrawn~\cite{Zavattini:2007ee}.

Laboratory experiments devised to search for an ALP-induced photon beam polarization suffer from the intrinsic limitation that a very short baseline is available. Remarkably enough, astrophysical observations of distant X-ray and ${\gamma}$-ray sources offer new opportunities to look for ALP effects in polarization measurements (see~\cite{Burrage:2008ii} for a recent comprehensive study). Actually, several satellite-borne missions are currently under consideration to perform the challenging measurement of the polarization state of photons emitted by distant astrophysical X-ray and $\gamma$-ray sources like the gamma-ray bursts (GRBs) in the keV-MeV energy range. These new missions have a wide field of view and a broad energy band, thereby guaranteeing polarimetric measurements of GRBs to be performed with high statistics. Such a circumstance evidently allows for a reliable determination of the statistical properties of the GRB polarization, which has recently been proposed as a crucial tool to discriminate among different models for the GRB emission~\cite{Toma:2008vi}. 

We show that strong modifications in the polarization pattern of distant X-ray and $\gamma$-ray sources can be produced by ALPs with parameters lying in experimentally allowed ranges. Basically, the following situation is envisaged. Photons are emitted by the considered sources at cosmological distances and on their way to us they cross different magnetic field configurations, which ought to induce   photon-ALP mixing. Consequently, the initial photon polarization gets changed and this change can eventually be detected. 

Manifestly, the role of the magnetic field is crucial in this respect and we have to contemplate all kinds of magnetic field configurations that the photon beam can experience, namely the magnetic field inside the source, the extragalactic magnetic field, the Galactic magnetic field and possibly an intracluster magnetic field if the beam crosses a cluster of galaxies. All these magnetic field components have a quite complicated and poorly known morphology, so that it has become customary to suppose that they possess a domain-like structure with varying coherence lengths. Therefore, we are actually dealing with a long baseline astrophysical setup to study ALP effects on the polarization of X-ray and ${\gamma}$-ray photons from distant sources. Our aim is indeed to investigate under which circumstances these effects are sizeable and detectable. For definiteness, our attention will be focussed on GRBs but our conclusions apply to 
any far-away astrophysical source of photons in the keV-MeV energy range, among which blazars play a very important role. We stress that at variance with a previous analysis of similar effects concerning radio sources~\cite{harari} our picture does not require the actual presence of an ALP cosmological background in the Universe but merely demands that ALPs are produced by photon oscillations.

The plan of the paper is as follows. In Section~2 we discuss the perspectives and the expectations for the measurement of GRB polarization. In Section~3 we review the mechanism of photon-ALP mixing in random magnetic fields which is relevant for our physical case and we identify the allowed ranges of the parameters that determine the effect in question. In Section~4 we discuss the impact of photon-ALP mixing on the polarization of GRBs for various intervening magnetic field configurations. We find that photon-ALP mixing in the magnetic field of GRBs is strongly suppressed. Still, during photon propagation from the source to us a sizeable mixing can occur. Due to the randomness of crossed magnetic fields, photon-ALP mixing depends strongly on the orientation of the line of sight.  As a consequence, this effect is expected to produce a peculiar broad distribution in the polarization of GRBs with different projected position on the sky. Thus, the detection of such a feature through future high-statistics polarimetric measurements can become a new strategy to look for the existence of ALPs with very low mass. Finally, in Section~5 we offer a discussion of our results as well as of the possible further developments in this field.

%%%%%%%%%%%%%%%%%%%%%%%%%%%%%%%%%%%%%%%%%%%%%%%%%%%%%%%%%%%
\section{Gamma-ray burst polarization measurements}
%%%%%%%%%%%%%%%%%%%%%%%%%%%%%%%%%%%%%%%%%%%%%%%%%%%%%%%%%%%
GRBs are brief, intense flashes of $\gamma$-rays originating at cosmological distances and they are the most luminous objects in the Universe. Presently available lightcurves and spectral observational information fail to provide a unique answer concerning their emission mechanism, and the polarization measurements of their $X$-ray and $\gamma$-ray emission have been recognized as a crucial mean to shed light on the inner structure of GRBs, 
including the geometry and the physical processes occurring close to the central engine. 

To date, GRB polarization measurements have been carried out systematically only in the optical band and in the afterglow phase. Sensitive observational techniques for 
$X$-ray and $\gamma$-ray polarimetry are being developed and in the next few years various polarimetric missions can finally enable to measure the $X$-ray and $\gamma$-ray polarization of the GRB emission. The missions presently under consideration include $POET$ (Polarimeters for Energetic Transients)~\cite{poet}, $PoGO$ (Polarimeter of Gamma-ray Observer)~\cite{Mizuno:2004ag}, $POLAR$~\cite{Produit:2005zu}, $GEMS$ (Gravity and Extreme Magnetism)~\cite{Jahoda:2007pd}, $XPOL$~\cite{Costa:2008pb}, $GRIPS$ (Gamma-Ray Burst Investigation via  Polarimetry and Spectroscopy)~\cite{Greiner:2008yd} and $NHXM$ (New Hard X-ray Mission)~\cite{Tagliaferri:2010wk}.

In particular, $POET$ is the only one that incorporates a broadband capability for measuring the prompt GRB emission and employs two different polarimeters, both with a wide field of view: $GRAPE$ (Gamma-Ray Polarimeter Experiment) operating between $60 \, {\rm keV}$ and $500 \, {\rm keV}$ 
and $LEP$ (Low Energy Polarimeter) operating between $2 \, {\rm keV}$ and $500 \, {\rm keV}$. Two smaller mission concepts based on the
$POET$ instruments are already scheduled for launch: $GRAPE$ and $GRBP$ (Gamma-Ray Burst Polarimeter, a smaller version of LEP).  
 Both instruments offer the opportunity for a first glimpse at the polarization of prompt GRB emission and in fact $POET$ provides sufficient sensitivity and sky coverage to detect up to 200 GRBs in a two-years mission.

In general, given a random distribution of viewing angles, a statistical study of the polarization properties of a large sample of GRBs should allow to 
discriminate among different emission models and can provide a direct diagnostic tool for the magnetic field structure, the radiation mechanism and 
the geometric configuration of GRB jets. Toma {\it et al}.~\cite{Toma:2008vi} display the predictions for the distribution of the amount of linear 
polarization in three different models: synchrotron emission with ordered magnetic fields (SO), synchrotron emission in random magnetic fields (SR) and Compton-drag model (CD). Specifically, the ratio $N_m/N_d$ of the number $N_m$ of GRBs for which the degree of polarization can be measured to the number $N_d$ of GRBs that are detected, and the distribution of the degree of linear polarization ${\Pi}_L$, can be used as criteria. It turns out that if $N_m/N_d>30\%$ and ${\Pi}_L$ clusters between 0.2 and 0.7, then the SO model will be favored. If instead $N_m/N_d<15\%$, then both the SR and the CD model will be preferred. Finally, if several events with ${\Pi}_L>0.8$ are observed, then the CD model will instead be singled out.

We will show that the presence of ALPs mixing with photons in astrophysical magnetic fields can drastically affect such expected statistical distributions for the linear polarization of GRBs.

%%%%%%%%%%%%%%%%%%%%%%%%%%%%%%%%%%%%%%%%%%%%%%%%%%%%%%%%%%%%%%%%%
\section{Photon-ALP mixing in random magnetic fields}
 %%%%%%%%%%%%%%%%%%%%%%%%%%%%%%%%%%%%%%%%%%%%%%%%%%%%%%%%%%%%%%%%%%%%%%%%%%%%%%%  

\subsection{Equations of motion}

The Lagrangian describing the photon-ALP system is 
\begin{equation}
{\cal L} = {\cal L}_\gamma +{\cal L}_a + {\cal L}_{a\gamma} \,\ .
\label{eq:lagr}
\end{equation}
The QED Lagrangian for  photons is
\begin{equation}
{\cal L_{\gamma}} = -\frac{1}{4} F_{\mu \nu} \, F^{\mu \nu} 
+ \frac{{\alpha}^2}{90 \, m^4_e} \, \left[ \left(F_{\mu \nu} \, F^{\mu \nu} 
\right)^2 + \frac{7}{4} \left(F_{\mu \nu} \, \tilde F^{\mu \nu} \right)^2 \right]~,
\label{eq:lagr211209}
\end{equation}
where $F_{\mu \nu} \equiv ({\bf E}, {\bf B})$ is the electromagnetic field tensor, $\tilde{F}_{\mu\nu} = \frac{1}{2}\epsilon_{\mu\nu\rho\sigma}F^{\rho\sigma}$ is its dual, $\alpha$ is the fine-structure constant and $m_e$ is the electron mass. Natural Lorentz-Heaviside units with with $\hbar=c=k_{\rm B}=1$ are employed throughout. The second term on the r.h.s. of Eq. (\ref{eq:lagr211209}) is the Euler-Heisenberg-Weisskopf (HEW) effective Lagrangian~\cite{Raffelt:1987im}, which accounts for the one-loop corrections to classical electrodynamics. 
%for photon frequencies $\omega \ll m_e$. 
The Lagrangian for the noninteracting ALP field $a$ is 
\begin{equation}
{\cal L}_a = \frac{1}{2} \, \partial^{\mu} a \, \partial_{\mu} a - \frac{1}{2} \, m^2 \, a^2 \,\ ,
\end{equation}
where $m$ is the ALP  mass. A general feature of ALP models is the CP-conserving pseudo-scalar two-photon coupling, so that the 
photon-ALP interaction is represented by the following Lagrangian~\cite{Raffelt:1987im}
\begin{equation}
{\cal L}_{a\gamma}=-\frac{1}{4} \,\gag
F_{\mu\nu}\tilde{F}^{\mu\nu}a=\gag \, {\bf E}\cdot{\bf B}\,a~,
\end{equation}
where $\gag$ is the photon-ALP coupling constant (which has the dimension of an inverse energy). It is always assumed ${\gag} \ll G_F^{1/2}$ and 
$m \ll G_F^{- 1/2}$, with  $G_F^{-1/2} \simeq 250 \, {\rm GeV}$ denoting the Fermi scale of weak interactions.

We shall be concerned throughout with a monochromatic photon/ALP beam of energy $E$ propagating along the $z$-direction in the presence of a magnetic field ${\bf B}$. Clearly, the beam propagation is described by the second-order coupled generalized Klein-Gordon and Maxwell equations arising from the Lagrangian
 in Eq.~(\ref{eq:lagr}). However, since we are interested in the regime $E \gg m$ the short-wavelength approximation  can be applied successfully and turns the beam propagation equation into the following Schr\"odinger-like one~\cite{Raffelt:1987im}
\begin{equation}
\label{we} 
\left(i \, \frac{d}{d z} + E +  {\cal M} \right)  \left(\begin{array}{c}A_x (z) \\ A_y (z) \\ a (z) \end{array}\right) = 0~,
\end{equation}
where $A_x (z)$ and $A_y (z)$ are the two photon linear polarization amplitudes along the $x$ and $y$ axis, respectively, $a (z)$ denotes the ALP amplitude  and ${\cal M}$ represents the photon-ALP mixing matrix. 

For our futher purposes, it is more convenient to work with the polarization density matrix 
\begin{equation}
\rho (z) = \left(\begin{array}{c}A_x (z) \\ A_y (z) \\ a (z)
\end{array}\right)
\otimes \left(\begin{array}{c}A_x (z)\  A_y (z)\ a (z)\end{array}\right)^{*}
\end{equation}
which obeys the Liouville-Von Neumann equation
\begin{equation}
\label{vne}
i \frac{d \rho}{d z} = [\rho, {\cal M}]
\end{equation}
associated with Eq.~(\ref{we}). We denote by $T(z,z_0)$ the transfer function, namely the solution of Eq.~(\ref{we}) with initial condition $T(z_0,z_0) = 1$. Then any solution of Eq.~(\ref{vne}) can be represented as
\begin{equation}
\label{vnea}
\rho (z) = T(z,z_0) \, \rho (z_0) \, T^{\dagger}(z,z_0)~.
\end{equation}

The mixing matrix ${\cal M}$ takes a simpler form if we restrict our attention to the case of a photon beam propagating in a single magnetic domain, where the magnetic field ${\bf B}$ is supposed to be homogeneous. We denote by ${\bf B}_T$ the transverse magnetic field, namely its component in the plane normal to the beam direction. We can choose the $y$-axis along ${\bf B}_T$ so that $B_x$ vanishes. Under these simplifying assumptions, the mixing matrix can be written as~\cite{Mirizzi:2006zy} 
\begin{equation}
{\cal M}^{(0)} =   \left(\begin{array}{ccc}
\Delta_{ \perp}  & 0 & 0 \\
0 &  \Delta_{ \parallel}  & \Delta_{a \gamma}  \\
0 & \Delta_{a \gamma} & \Delta_a 
\end{array}\right)~,
\label{eq:massgen}
\end{equation}
whose elements are~\cite{Raffelt:1987im}
\begin{equation}
\Delta_\parallel \equiv \Delta_{\rm pl} + 3.5 \, \Delta_{\rm QED}~,
\end{equation}
\begin{equation}
\Delta_\perp \equiv \Delta_{\rm pl} + 2 \, \Delta_{\rm QED}~,
\end{equation}
\begin{equation}
\Delta_{a\gamma} \equiv \frac{1}{2} g_{a\gamma} B_T \simeq 1.52 \times 10^{-2} \left(\frac{g_{a\gamma}}{10^{-11} \, {\rm GeV}^{-1}} \right)
\left(\frac{B_T}{10^{-9} \,\rm G}\right) {\rm Mpc}^{-1}~,
\end{equation}
\begin{equation}
\Delta_a \equiv - \frac{m^2}{2E} \simeq -7.8 \times 10^{-3} \left(\frac{m}{10^{-13} \, {\rm eV}}\right)^2 \left(\frac{E}{{10^2 \, \rm keV}} \right)^{-1} 
{\rm Mpc}^{-1}~,
\end{equation}
with
\begin{equation}
\Delta_{\rm pl} \equiv -\frac{\omega^2_{\rm pl}}{2E} \simeq - 1.1 \times10^{-4} \left(\frac{E}{{10^2\, \rm keV}}\right)^{-1} \left(\frac{n_e}{10^{-7} \, {\rm cm}^{-3}}\right) {\rm Mpc}^{-1}~,
\end{equation}
\begin{equation}
\Delta_{\rm QED} \equiv \frac{\alpha E}{45 \pi} \left(\frac{B_T}{B_{\rm cr}} \right)^2 \simeq  4.1 \times 10^{-16}\left(\frac{E}{{10^2\, \rm keV}}\right)
\left(\frac{B_T}{10^{-9}\,\rm G}\right)^2 {\rm Mpc}^{-1}~,
\end{equation}
where $n_e$ is the electron density in the medium, $\omega^2_{\rm pl} = 4 \pi \alpha n_e/m_e$ is the associated plasma frequency, $B_{\rm cr} \equiv m^2_e /e \simeq 4.41 \times 10^{13} \, {\rm G}$ is the critical magnetic field and $e$ denotes the electron charge.

%%%%%%%%%%%%%%%%%%%%%%%%%%%%%%%%%%%%%%%%%%%%%%%%%%%%%%%%%%%%%%%%%% 
\subsection{Input parameters} 
%%%%%%%%%%%%%%%%%%%%%%%%%%%%%%%%%%%%%%%%%%%%%%%%%%%%%%%%%%%%%%%%%%%% 

Before proceeding further, we find it convenient to discuss the ranges of the parameters entering our scenario. 
 The strength of the widespread, all-pervading magnetic field in the extragalactic medium has to meet the constraint $B \lesssim2.8\times10^{-7} (L/{\rm Mpc})^{-1/2}\,{\rm G}$ -- where $L$ denotes its coherence length -- which arises by scaling the original bound from the Faraday effect of distant
radio sources~\cite{Kronberg:1993vk,Grasso:2000wj} to the now much better known baryon density measured by the Wilkinson Microwave Anisotropy Probe (WMAP) mission~\cite{Hinshaw:2008kr}. Its coherence length is expected to lie in the range $1 \, {\rm Mpc} < L < 10 \, {\rm Mpc}$~\cite{Blasi:1999hu}. The mean diffuse intergalactic plasma density is bounded by $n_e \lesssim 2.7 \times 10^{-7}$~cm$^{-3}$, corresponding to the WMAP measurement of the baryon density~\cite{Hinshaw:2008kr}. Recent results from the CAST experiment yield a direct bound on the photon-ALP coupling constant $\gag\lesssim 8.8\times 10^{-11}$~GeV$^{-1}$ for $m \lesssim 0.02$~eV~\cite{Arik:2008mq}, 
slightly better than the long-standing globular-cluster limit~\cite{Raffelt:2006cw}. In addition, for $m \lesssim 10^{- 10}$~eV a more stringent limit arises from the absence of $\gamma$-rays from SN~1987A, giving $\gag\lesssim 1\times 10^{-11}$~GeV$^{-1}$~\cite{Brockway:1996yr,Grifols:1996id} 
even if with a large uncertainty\footnote{These bounds on ALPs can be relaxed if they have a chameleontic nature~\cite{Brax:2007ak}, in which case the most stringent constraint comes from the observed starlight polarization and reads $g_{a \gamma} \lesssim 10^{-9}$~GeV$^{-1}$\cite{Burrage:2008ii}. However, we do not commit ourselves to chameleontic ALPs.}.

%%%%%%%%%%%%%%%%%%%%%%%%%%%%%%%%%%%%%%%%%%%%%%%%%%%%%%%%%%%%%%%%%% 
\subsection{Mixing in a single magnetic domain} 
%%%%%%%%%%%%%%%%%%%%%%%%%%%%%%%%%%%%%%%%%%%%%%%%%%%%%%%%%%%%%%%%%%%% 
 
Inside a single magnetic domain ${\bf B}$ is homogeneous. Then 
it is straightforward to check that ${\cal M}^{(0)}$ in Eq. (\ref{eq:massgen}) can be brought into a diagonal form 
\ben  \label{rot_to_prim}
D=
\left(
\begin{array}{ccccccccc}
D_1&0&0\\
0&D_2&0\\
0&0&D_3
\end{array}
\right)
\end{equation}
by the similarity transformation 
\begin{equation}
D = W \, {\cal M}^{(0)} \, W^{\dagger}
\end{equation}
with 
\ben  \label{rot_to_prim}
W=
\left(
\begin{array}{ccccccccc}
1&0&0\\
0&\cos\theta&\sin\theta\\
0&-\sin\theta&\cos\theta
\end{array}
\right)~,
\end{equation}
where the mixing angle $\theta$ is given by
\ben  \label{theta}
\theta = \frac{1}{2} \arctan \left(\frac{ 2 \Delta_{a \gamma}}{\Delta_{\parallel}-\Delta_a} \right)
\een
and explicitly we get
\begin{equation}
D_{1} = \Delta_{\perp}~,
\end{equation}
\begin{equation}
D_{2} = \frac{(\Delta_{\parallel} + \Delta_{a})}{2} + \frac{1}{2} \left[(\Delta_{a} - \Delta_{\parallel})^2 + 4\Delta_{a \gamma}^{2}\right]^{1/2}~, 
\end{equation}
\begin{equation}
D_{3} =  \frac{(\Delta_{\parallel} + \Delta_{a})}{2} - \frac{1}{2} \left[(\Delta_{a} - \Delta_{\parallel})^2 + 4\Delta_{a \gamma}^{2}\right]^{1/2}~.
\end{equation}
So, we find that the transfer matrix is presently 
\begin{equation}
\label{mra}
T^{(0)} (z,z_0) = W^{\dagger} \, e^{i D (z - z_0)} \, W~.
\end{equation}
Finally, the  density matrix can be computed by means of the general formula in Eq.~(\ref{vnea}).

The probability that a photon initially polarized along the $y$ axis ($\rho_{22} (0)=1$) converts into an ALP after a distance $d$ then reads~\cite{Raffelt:1987im}
\begin{equation}
\label{a16}
P^{(0)}_{a \gamma} = \rho_{33} (d) ={\rm sin}^2 2 \theta \  {\rm sin}^2
\left( \frac{\Delta_{\rm osc} \, d}{2} \right)~,
\end{equation}
where the oscillation wave number is
\begin{equation}
\label{a17}
{\Delta}_{\rm osc} = \left[\left( \Delta_a - \Delta_{\parallel} \right)^2 + 4 \Delta_{a \gamma}^2 \right]^{1/2}~.
\end{equation}
Experience with problems similar to the one considered here shows that it proves useful to define a {\it low critical energy}~\cite{De Angelis:2007yu}
\begin{equation}
E_L \equiv  \frac{E \, |\Delta_a- \Delta_{\rm pl}|}{2 \, \Delta_{a \gamma}}
\simeq  \frac{25 \, | m^2 - {\omega}_{\rm pl}^2|}{(10^{-13}{\rm eV})^2}
\left( \frac{10^{-9}{\rm G}}{B_T} \right) \left( \frac{10^{-11}\rm GeV^{-1}}{g_{a \gamma}} \right) {\rm keV} 
\label{eq:EL}
\end{equation}
along with a {\it high critical energy}~\cite{Bassan:2009gy}
\begin{equation}
E_H \equiv \frac{90 \pi \,g_{a \gamma} \, B^2_{\rm cr}}{7 \alpha \,B_T} \simeq 2.1\times 10^{15} \left(\frac{10^{-9} \, \rm{G}}{B_T}\right) 
\left(\frac{g_{a \gamma}}{10^{-11}\rm GeV^{-1}}\right) {\rm keV}~.
\label{eq:eH}
\end{equation}
It is easy to see that the oscillation wave number can be expressed in terms of $E_L $ and $E_H$ as
\begin{equation}
\label{a17t}
{\Delta}_{\rm osc} = 2 \Delta_{a \gamma} \left\{1+ \left[ \textrm{sgn}(m_a - \omega_{\rm pl}) \left(\frac{E_L}{E} \right) + \left(\frac{E}{E_H} \right) \right]^2 \right\}^{1/2}
\end{equation}
and from Eqs. (\ref{theta}), (\ref{a16}), (\ref{a17}) and (\ref{a17t}) it follows that 
in the energy range $ E_L\ll E \ll E_H$  
 the  photon-ALP mixing is maximal ($\theta \simeq \pi/4$) and the conversion probability becomes energy-independent. This is the so-called {\it strong-mixing regime}. Outside this regime the conversion probability turns out to be energy-dependent and vanishingly small, so that $E_L$ and $E_H$ acquire the meaning of low-energy and high-energy energy cut-off, respectively.

 %%%%%%%%%%%%%%%%%%%%%%%%%%%%%%%%%%%%%%%%%%%%%%%%%%%%%%%%%%%%%%%5
 \subsection{Photon transfer function and polarization}
 %%%%%%%%%%%%%%%%%%%%%%%%%%%%%%%%%%%%%%%%%%%%%%%%%%%%%%%%%%%
As stressed in Section 1, the magnetic field crossed by the beam on its way from the source to us is modelled as a network of magnetic domains. For 
each magnetic field component (source, extragalactic,  intracluster and Galactic) all domains are supposed to have the same size $L$ equal to the coherence length and in every domain the  magnetic field ${\bf B}$ is assumed to have the same strength but its direction is allowed to change randomly from one domain to another. 

As a matter of fact, the application of this approach to extragalactic magnetic fields neglects cosmological effects which are potentially important, since the GRB polarization can be measured for sources out to a redshift of about 2\footnote{Private communication from E. Costa and K. Toma.}.  However, for the sake of clarity -- and even because this effect is irrelevant outside cosmology -- we present below the investigation of the photon-ALP mixing in the above random network discarding redshift-dependent complications, which will be discussed in Subsection 3.5.

In order to accomplish  our task, we take the $x,y,z$ coordinate system as {\it fixed} once and for all, denoting by $\psi$ the angle between ${\bf B}_T$ and the $y$ axis in a generic domain and treating $\psi$ as a {\it random variable} in the range $0 \leq \psi < 2 \pi$. 

So, what we need in the first place is the generalization of the previous result for an arbitrary orientation of ${\bf B}_T$ in a single domain. Clearly, the mixing matrix ${\cal M}$ presently arises from ${\cal M}^{(0)}$ through the similarity transformation
\begin{equation}
\label{mr101109}
{\cal M}=V^{\dagger} (\psi) \, {\cal M}^{(0)} \, V (\psi) 
\end{equation}
operated by the rotation matrix in the $x$-$y$ plane, namely 
\begin{equation} 
V (\psi) =
\left(
\begin{array}{ccccccccc}
\cos\psi&-\sin\psi&0\\
\sin\psi&\cos\psi&0\\
0&0&1
\end{array}
\right) \,\ .
\end{equation}
Accordingly we find~\cite{Mirizzi:2006zy} 
\begin{equation}
\label{aa8MR}
{\cal M} = \left(
\begin{array}{ccc}
\Delta_{xx} & \Delta_{xy} & \Delta_{a\gamma} \, \sin\psi\\
\Delta_{yx} & \Delta_{yy} & \Delta_{a\gamma} \, \cos\psi\\
\Delta_{a\gamma} \, \sin\psi& \Delta_{a\gamma} \, \cos\psi& \Delta_{a} \\
\end{array}
\right)~,
\end{equation} 
with 
\begin{equation}
\Delta_{xx} = \Delta_\parallel \, \sin^2 \psi + \Delta_\perp \cos^2 \psi~,
\end{equation}
\begin{equation}
\Delta_{xy} = \Delta_{yx}=(\Delta_\parallel -\Delta_\perp) \sin\psi \, \cos\psi~,
\end{equation}
\begin{equation}
\Delta_{yy} = \Delta_\parallel \cos^2 \psi + \Delta_\perp \sin^2 \psi~.
\end{equation}

Our next step consists in the evaluation of the transfer matrix in the considered domain. As before, our strategy is to diagonalize the mixing matrix. Dealing with 
${\cal M}$ is slightly more complicated than dealing with ${\cal M}^{(0)}$, but Eq.~(\ref{mr101109}) allows to reduce the present problem to the one solved above~\cite{Christensson:2002ig}. 
It is convenient to label all quantities pertaining to the generic $k$-th domain with the index $k$, with the understanding that all $z$-dependent quantities labelled by $k$ are evaluated at the edge of the $k$-th domain closer to us. The source is located at $z=0$ and we suppose that there are 
$N$ magnetic domains with size $L$ along the line of sight. Hence, the beam propagation over the $k$-th domain is described by the polarization density matrix
\begin{equation}
\rho_k = T_k \, \rho_{k-1} \, T^{\dagger}_{k}~,
\end{equation}
where the transfer matrix $T_k$ can be rewritten as
\begin{equation}
T_k = V^{\dagger} (\psi_k) \, W^{\dagger} \, e^{i D L} \, W \, V (\psi_k)~,
\end{equation}
where $\psi_k$ is the random value of $\psi$ in the domain in question. The explicit form of $T_k$ can be represented as
\begin{equation}
T_k = e^{i D_1 L} \, T_A (\psi_k) + e^{i D_2 L} \, T_B (\psi_k) + e^{i D_3 L} \, T_C (\psi_k)~, 
\end{equation}
with
\begin{equation}
T_A (\psi_k) \equiv
\left(
\begin{array}{ccc}
\cos^2 \psi_k & -\sin \psi_k \cos \psi_k & 0 \\
- \sin \psi_k \cos \psi_k & \sin^2 \psi_k & 0 \\
0 & 0 & 0
\end{array}
 \right)~,
\end{equation}
\begin{equation} 
T_B (\psi_k) \equiv
\left(
\begin{array}{ccc}
\sin^2 \psi_k \cos^2 \theta & \sin \psi_k \cos \psi_k \cos^2 \theta & 
\sin \theta \cos \theta \sin \psi_k \\ 
\sin \psi_k \cos \psi_k \cos^2 \theta & \cos^2 \psi_k \cos^2 \theta & 
\sin \theta \cos \theta \cos \psi_k \\
\sin \psi_k \cos \theta \sin \theta & \cos \psi_k \sin \theta \cos \theta & \sin^2 \theta 
\end{array}
\right)~,
\end{equation} 
\begin{equation}
T_C (\psi_k) \equiv 
\left(
\begin{array}{ccc}
\sin^2 \theta \sin^2 \psi_k & \sin^2 \theta \sin \psi_k \cos \psi_k & - \sin \theta \cos \theta \sin \psi_k \\
\sin^2 \theta \sin \psi_k \cos \psi_k & \sin^2 \theta \cos^2 \psi_k & - \sin \theta \cos \theta \cos \psi_k \\
- \sin \theta \cos \theta \sin \psi_k & - \sin \theta \cos \theta \cos \psi_k & \cos^2 \theta 
\end{array}
 \right)~.
\end{equation}

Altogether, the whole beam propagation from the source to us is described by
\begin{equation}
\rho_N = T (\psi_N, ... , \psi_1) \, \rho_0 \, T^{\dagger} (\psi_N, ... , \psi_1)~,
\end{equation}
where $\rho_0$ is the polarization density matrix at emission and we have set
\begin{equation}
T (\psi_N, ... , \psi_1) \equiv \prod^N_{k = 1} T_k~.
\end{equation}

Finally, the $2 \times 2$ photon polarizaton density matrix (i.e. the 1-2 block of the density matrix
for the photon-ALP system) can be expressed in terms of the Stokes parameters as~\cite{kosowski}
\begin{equation}
%\label{aa8MR}
{\rho}_{\gamma} = \frac{1}{2} \left(
\begin{array}{cc}
I + Q & U - i V \\
U + i V & I - Q\\
\end{array}
\right)
\end{equation} 
and the degree of {\it linear polarization} ${\Pi}_L$ is defined as~\cite{rybicki}
\begin{equation}
{\Pi}_L  \equiv \frac{(Q^2 + U^2)^{1/2}}{I} =
 \frac{ \left[ \left( \rho_{11} - \rho_{22} \right)^2 + \left( \rho_{12} - \rho_{21} \right)^2 \right]^{1/2}}{ \rho_{11} + \rho_{22} }~.
\end{equation}

In the following, we will evaluate the ALP contribution to ${\Pi}_L$ in the various physical situations mentioned above.

%..............................................................
\label{sec:universe}
\subsection{Effects of the cosmic expansion}
%............................................................
As explained in the previous Subsection 3.4, the considered strategy has to be slightly revised in order to account for the cosmic expansion when considering photon propagation in the extragalactic magnetic fields. In practice, the logic remains the same but distances have to be parameterized in term of the redshift $z$ rather than by the coordinate along the propagation direction\footnote{In the present Subsection, $z$ denotes from now on the redshift rather than the $z$-coordinate, and so no confusion will arise.}. As is well known, the distance $dl(z)$ travelled by a photon over an infinitesimal redshift interval $dz$ is expressed by the relation
\begin{equation}
\label{lungh}
d l(z) = \frac{dz}{H_0 \, \left(1 + z \right) \left[ \left(1 + z \right)^2 \left({\Omega}_M z + 1  \right) - {\Omega}_{\Lambda} z \left(z + 2 \right) \right]^{1/2}}~,
\end{equation}
where for the standard $\Lambda$CDM cosmological model we have $H_0 \simeq 72 \, {\rm Km} \, {\rm s}^{-1} \, {\rm Mpc}^{-1}$ for the Hubble constant, while ${\Omega}_M \simeq 0.3$ and ${\Omega}_{\Lambda} \simeq 0.7$ represent the average cosmic density of matter and dark energy, respectively, in units of the critical density 
${\rho}_{\rm cr} \simeq 0.92 \cdot 10^{- 29} \, {\rm g} \, {\rm cm}^{- 3}$~\cite{Amsler:2008zzb}. 
Hence, a generic domain extending over the redshift interval $[z_a,z_b]$ ($z_a < z_b$) has size
\begin{equation}
\label{lunghG}
L (z_a,z_b) \simeq 4.17 \int_{z_a}^{z_b} \frac{d z}{\left(1 + z \right) \left[ \left(1 + z \right)^2 \left(0.3 z + 1  \right) - 0.7 z \left(z + 2 \right) \right]^{1/2}} \, {\rm Gpc}~.
\end{equation}

After photons have been emitted by a GRB, they propagate in the intergalactic medium (IGM). Within the present context, it looks natural to describe the overall structure of the cellular configuration of the extragalactic magnetic field by a {\it uniform}  mesh in redshift space, which can be constructed as follows. In the lack of any reliable information, we assume for definitiveness 
 $L_0=1$~Mpc as  coherence length of the magnetic field $B$ at redshift $z = 0$, which in our notation translates into 
\begin{equation}
\label{mag2Z}
L (0,z_1) =  1 \, {\rm Mpc}~, 
\end{equation}
which fixes the size in redshift space of the domain closest to us through Eq. (\ref{lunghG}). Actually, since we are employing a uniform mesh in redshift space, $z_1$ sets the size of {\it all} magnetic domains by recursive application of Eq. (\ref{lunghG}). Hence, the $n$-th one extends from $z = (n - 1) z_1$ to 
$z= n z_1$ and its length can be written as $L ((n - 1) z_1, n z_1)$. 

The absence of the Gunn-Peterson effect is usually taken as evidence that the IGM is ionized. The high electrical conductivity of the IGM allows us to assume that the electron number density $n_e$ traces the cosmic matter distribution. But since $n_e$ is proportional to the mass density $\rho$ and the average number density of electrons $\bar n_e$ goes like $(1 + z)^3$, we obtain 
relationship
\begin{equation}
n_e(z) = (1 + \delta(z)) {\bar n}_{e,0} (1 + z)^3  \,\ ,
\end{equation}
where 
$\delta(z) \equiv (\rho(z)-\bar\rho(z)) / \bar\rho(z)$ is the mass density contrast, which quantifies the local deviation of the matter density from the average and can be computed starting from a given spectrum of density perturbations (see, e.g.~\cite{Christensson:2002ig}). Actually, two facts follow at once. First, the ionized nature of the IGM implies that the plasma frequency varies with $z$ as
\begin{equation}
\label{mag2}
{\omega}_{\rm pl} = (1 + \delta(z))^{1/2} {\bar\omega}_{{\rm pl},0} \, \left(1 + z \right)^{3/2}
\end{equation}
owing to its dependence on $n_e$ (${\bar\omega}_{{\rm pl},0}$ obviously corresponds to ${\bar n}_{e,0}$). Second, because of the resulting high conductivity of the IGM, the magnetic flux lines can be thought as frozen in the IGM. Therefore, flux conservation during the cosmic expansion entails that $B$ scales like  $n_e^{2/3}$, thereby implying
the magnetic field in the domain at redshift $z$ is
\begin{equation}
\label{mag1}
B = (1 + \delta(z))^{2/3} \,  {\bar B}_0 \left(1 + z \right)^2~,
\end{equation}
where ${\bar B}_0$ denotes the average magnetic field at $z=0$.

As a rule, simulations of the mass distribution in the Universe predict that most of the considered domains should be under-dense with respect to the mean by a factor $\sim 10$~\cite{Csaki:2001jk,Valageas:1999wa}. Simulations as well as observations also show the existence of over-dense regions with respect to the mean by a factor $10-–100$ at low redshift~\cite{Dave:1998gm}, and Eq.~(\ref{mag1}) entails that in the latter regions the magnetic field should be larger by a factor $\sim 5$--$20$. However, the characteristic size of such over-dense regions turns out to be $\lesssim 100$~kpc, and so they fill only a small fraction of space~\cite{Schaye:2001me}. As a consequence, the overall effect of these over-dense regions on photon-ALP mixing is negligible: ignoring them effectively amounts to take the source slight nearer and nothing else. We are therefore led for simplicity to discard matter overdensity effects, thereby setting $\delta(z)=0$. In this way, we also avoid committing ourselves with any specific (model-dependent) choice for the matter density distribution. Hence, in the following we will simply assume $B \simeq {\bar B}_0 \left(1 + z \right)^2$, where  for definitiveness we will take $\bar B_0=10^{-9}$~G as average value of the magnetic field at $z=0$.  

Moreover, the exact value of ${\omega}_{\rm pl}$ and its scaling with the redshift [Eq.~(\ref{mag2})] can also be safely neglected. The reason is that we are working within the strong-mixing regime, in which plasma effects are irrelevant, and so our result is independent of the precise value of ${\omega}_{\rm pl}$ in every magnetic domain: all what matters is that the strong-mixing condition is met in any domain and we have cheked that for our preferred values of the input parameters this is indeed the case. Finally, we remark that another consequence of the strong-mixing regime is that photon-ALP mixing is energy-independent, and so we do not have to worry about the change of the photon energy along the line of sight caused by the cosmic expansion.

%----------------------------------------------
\section{ALP effects on the polarization of gamma-ray sources}
%----------------------------------------------

Equipped with the results of the previous Section, we are now ready to investigate the implications of photon-ALP mixing on the polarization
of GRBs. Needless to say, we also have to keep the photon-ALP conversion probability under control in order to make sure that the resulting  dimming does not prevent the source from being detected. As far as the magnetic field is concerned, we will address in turn the various configurations crossed by the beam, namely the component inside the source, the extragalactic and the intracluster contributions and the Galactic component. For definiteness, we shall take $g_{a \gamma} = 10^{- 11} \, {\rm GeV}^{- 1}$ throughout.

%%%%%%%%%%%%%%%%%%%%%%%%%%%%%%%%%%%%%%%%%%%%%%%%%%%%%%%%%%%%%%%%%%%%%%%%
\subsection{Source magnetic field} 
%%%%%%%%%%%%%%%%%%%%%%%%%%%%%%%%%%%%%%%%%%%%%%%%%%%%%%%%%%%%%%%%%%%%%%
The relevance of ALP-induced polarization in the source magnetic field for the GRB emission has been studied in~\cite{Rubbia:2007hf}. In that paper, the authors claim a potentially large effect. Lacking a detailed model for the  GRB environment, we follow~\cite{Rubbia:2007hf} and we consider the source as a region where the magnetic field is homogeneous, with strength $B \simeq 10^9$~G and size $L \simeq 10^9$~cm. The typical electron number density in the source is estimated as $n_e \simeq 10^{10}$~cm$^{-3}$. Accordingly, from Eq.~(\ref{eq:eH}) we find the the high-energy cut-off (arising  from QED vacuum polarization) is $E_H\simeq 2 \times 10^{-3}$~keV, regardless of $m$. Thus {\bf --} contrary to the previous claim {\bf --} in this  situation ALPs do not affect the GRB polarization. However, this model for the source is oversimplified and in particular the magnetic field could have a turbulent and irregular structure.
Since there are large uncertainties in the parameters of the GRB emitting regions,
it would be  premature to exclude definitively the possibility of any effect of ALP conversions in the source.
However, lacking of a detailed model for the GRB emitting region, hereafter we neglect  possible photon-ALP conversions occurring in the source.

%.............................................................
\subsection{Extragalactic magnetic field}
%..........................................................

Once emitted by a GRB, photons propagate in the IGM, which is modelled as explained in Subsection 3.5. Our results can be summarized as follows.

In Figure~1 we show the average final linear polarization ${\Pi}_L$ as a function of the photon energy $E$, starting with a completely unpolarized source at distance $d=100$~Mpc. The averaging process is over an ensemble of $10^4$ realizations of the random magnetic field network. Since in this case we are considering a relatively close source, we neglect the redshift dependence of the extragalactic magnetic field. In agreement with Eq.~(\ref{eq:EL}), we find that ALP effects on the polarization start to become relevant at $E\simeq 25$~keV for $m=10^{-13}$~eV and at $E\simeq 2.5$~keV for $m=10^{-14}$~eV. Moreover, ALP effects saturate fast as we enter the strong-mixing regime, where the photon-ALP conversion probability is energy-independent. A similar trend concerns the photon survival probability $P_{\gamma\gamma}$. In the following, we will focus on the strong-mixing regime, choosing $E=100$~keV as observed reference photon energy and $m=10^{-13}$~eV as reference ALP mass. 

In Figure~2 we exhibit the average linear polarization ${\Pi}_L$ and the photon survival probability $P_{\gamma\gamma}$ 
as a function of the source distance $d$, starting from a source fully polarized along the $x$ direction ($\rho_{xx} (0)=1$). Again, the averaging process is over an ensemble of $10^4$ realizations of the random magnetic field network. For our input values, we get $\Delta_{a\gamma} \, L\ll 1$ in every magnetic domain, so that the conversion probability in a single domain dictated by Eq.~(\ref{a16}) assumes the form $P^{(0)}_{a\gamma} = (\Delta_{a\gamma} \, L)^2$. In this limit, neglecting the redshift dependence in the extragalactic magnetic field, the average photon survival probability reduces to the simple analytic expression~\cite{Mirizzi:2006zy,MirizziHigh}
%..........................................................................
\begin{equation}
P_{\gamma\gamma}(d) = \frac{2}{3}+ \frac{1}{3}\exp\left(-\frac{3 P^{(0)}_{a\gamma} d}{L} \right) \,\ ,
\end{equation}
%......................................................................
thereby implying on average a complete equipartition of the photon/ALP ensemble after passage through many domains ($d\gg L$). As a check of our numerical simulations, in Figure~2 we also compare the numerical (dashed curve) and the analytical (continuous curve) results for $P_{\gamma\gamma}(d)$. The nice agreement between the two prescriptions reassures us about the reliability of our simulations. We find that the average photon survival probability saturates at $2/3$ for $d\gtrsim 500$~Mpc. Instead, the saturation of the average final linear polarization occurs at a larger distance $d \gtrsim 10^{3}$~Mpc with asymptotic value ${\Pi}_L \simeq 0.79$.  Furthermore, an important conclusion to be drawn from Figure~2 is that the dimming arising from photon-ALP conversion never prevents the observation of the gamma-ray flux.

%%%%%%%%%%%%%%%%%%%%%%%%%%% FIGURE 1 %%%%%%%%%%%%%%%%%%%%%%%%%%%%%%%%%
\begin{figure}[t]
\centering
\epsfig{figure=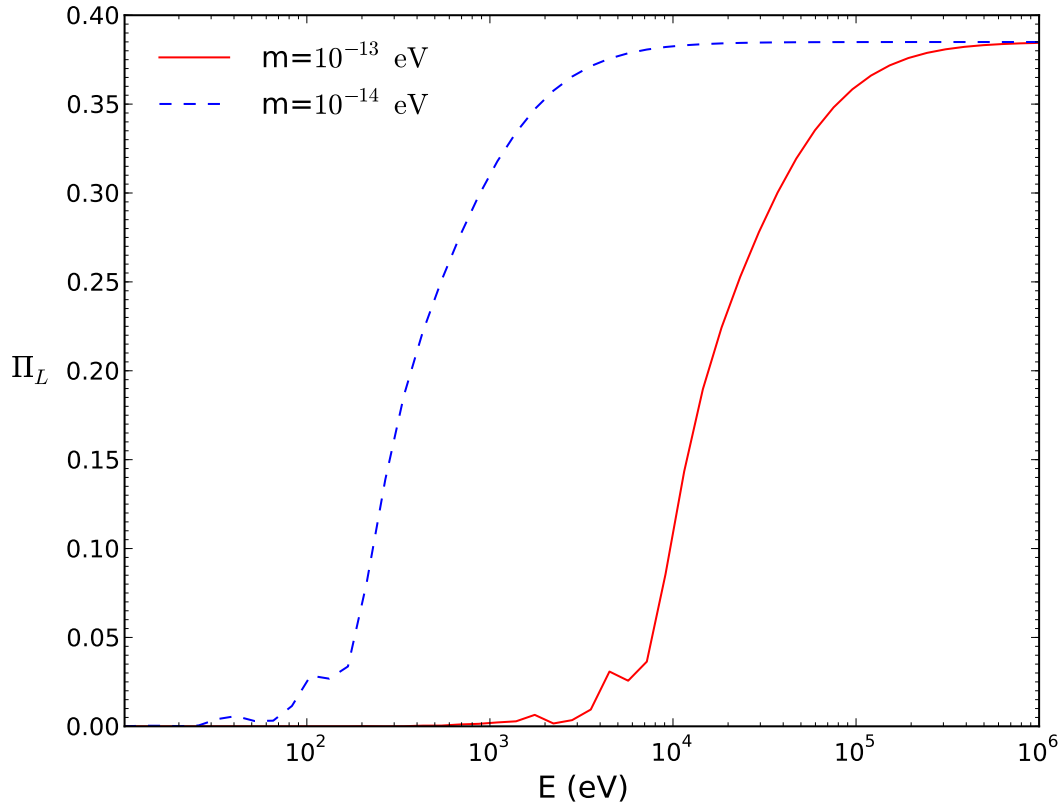, width =1.\columnwidth,  angle=0}
 \caption{
Average final linear polarization ${\Pi}_L$ as a function of the photon energy $E$ after propagation in the extragalactic magnetic field for ALP 
masses $m=10^{-13}$~eV (solid line) and $m=10^{-14}$~eV (dashed line), respectively. The emitting GRB is assumed to be completely unpolarized and at distance $d=100$~Mpc.
\label{fig1}}
\end{figure}
%%%%%%%%%%%%%%%%%%%%%%%%%%%%%%%%%%%%%%%%%%%%%%%%%%%%%%%%%%%%%%%%%%%%%

%%%%%%%%%%%%%%%%%%%%%%%%%%% FIGURE 2 %%%%%%%%%%%%%%%%%%%%%%%%%%%%%%%%%
\begin{figure}[t]
\centering
\epsfig{figure=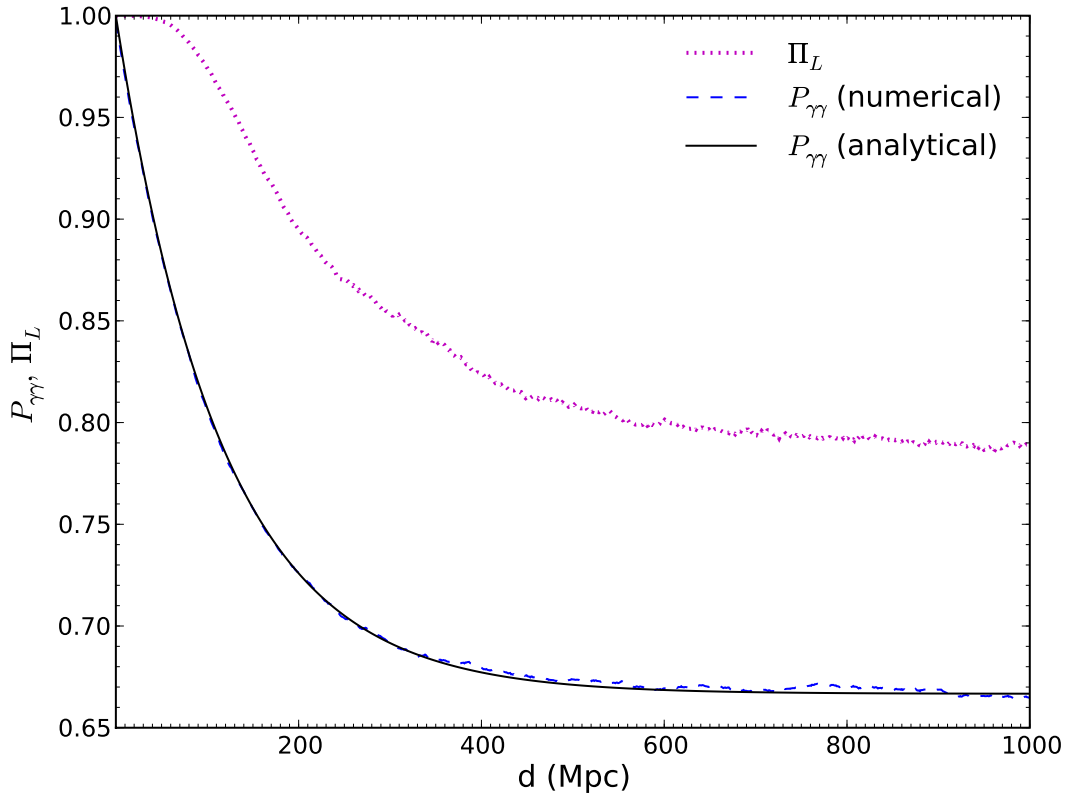, width =1.\columnwidth,  angle=0}
 \caption{
Average  linear polarization ${\Pi}_L$ (dotted line) and photon survival probability $P_{\gamma \gamma}$ evaluated numerically (dashed line) and analytically (solid line), as a function of the source distance $d$ for photon-ALP mixing in the extragalactic magnetic field. The source is assumed to be fully polarized 
along the $x$ direction.
\label{fig2}}
\end{figure}
%%%%%%%%%%%%%%%%%%%%%%%%%%%%%%%%%%%%%%%%%%%%%%%%%%%%%%%%%%%%%%%%%%%%% 

%%%%%%%%%%%%%%%%%%%%%%%%%%% FIGURE 3 %%%%%%%%%%%%%%%%%%%%%%%%%%%%%%%%%
\begin{figure}[t]
\centering
\epsfig{figure=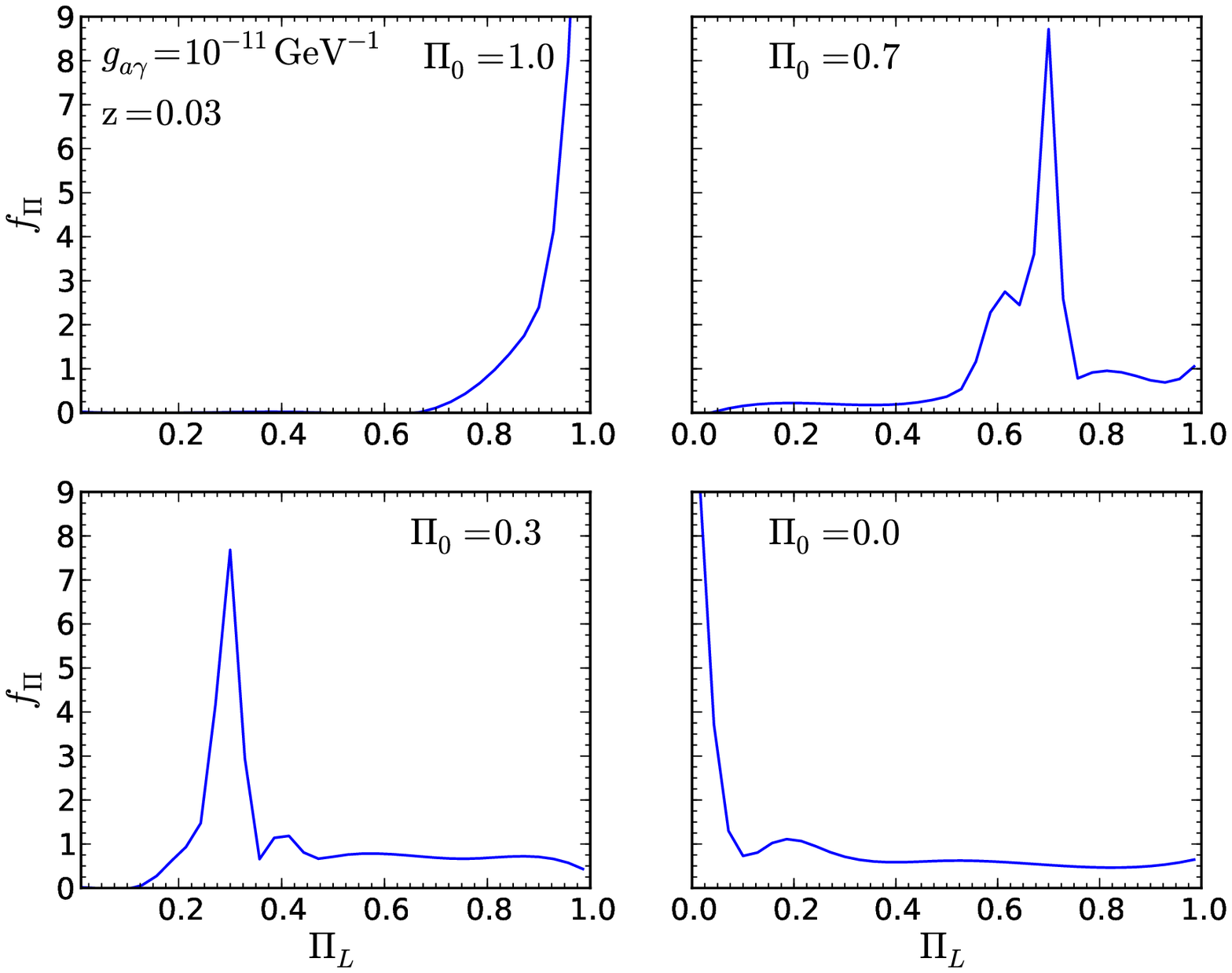, width =1.\columnwidth,  angle=0}
 \caption{
Probability density function $f_\Pi$ for the final linear polarization ${\Pi}_L$ after propagation in the extragalactic magnetic field, considering $10^4$ GRBs at redshift $z=0.03$, with initial linear polarization $\Pi_0=0.0,0.3,0.7,1.0$.
\label{fig3}}
\end{figure}
%%%%%%%%%%%%%%%%%%%%%%%%%%%%%%%%%%%%%%%%%%%%%%%%%%%%%%%%%%%%%%%%%%%%% 

In Figures~3, 4, 5 and 6 we display the final probability density functions $f_\Pi$ for the GRB linear polarization, assuming that all sources are at redshift $z=0.03$, $z=0.3$, $z=1$ and $z=2$, respectively.
From our Eq.~(\ref{lunghG}), these redshifts would correspond to distances $d \simeq 122.2$~Mpc, $d \simeq 1.2$~Gpc, 
$d \simeq 2.4$~Gpc, $d \simeq 3.5$~Gpc, respectively.
 In these simulations we are taking into account the redshift effects discussed in Subsection 3.5. To obtain the probability distributions, we have performed $10^4$ simulations of the photon evolution within the random magnetic field configurations. For simplicity, we have assumed that in each simulation all sources have the same initial linear polarization, $\Pi_0=0.0,0.3,0.7,1.0$, rispectively. It turns out that photon-ALP mixing smears out the initial GRB linear polarization. This effect increases with the source distance and saturates at $z\simeq 2$. For $z\lesssim 1$, $f_\Pi$ still presents a peak which is the relic of the initial linear polarization, but with long tails which are not present in the standard polarization distributions described in Section 2. In particular, the expected clustering in the linear polarization distributions gets smeared out by photon-ALP mixing. Moreover, since the standard GRB linear polarization should have an initial spread and the GRBs are actually distributed over a variety of distances, in the presence of photon-ALP mixing the final distributions should to be even more smeared out and irregular than the ones presented in these Figures. For $z\gtrsim 1$ we find rather 
flat probability distributions in the GRB polarizations, without any record of the initial linear polarization $\Pi_0$. As a consequence, the presence of photons-ALP mixing appears to hinder the possibility to extract from observational data information on the initial polarization mechanism for GRBs. Conversely -- in the lack of any standard explanation -- detection of the features presented above can be interpreted as a hint at the existence of very light ALPs.

%%%%%%%%%%%%%%%%%%%%%%%%%%% FIGURE 3 %%%%%%%%%%%%%%%%%%%%%%%%%%%%%%%%%
\begin{figure}[t]
\centering
\epsfig{figure=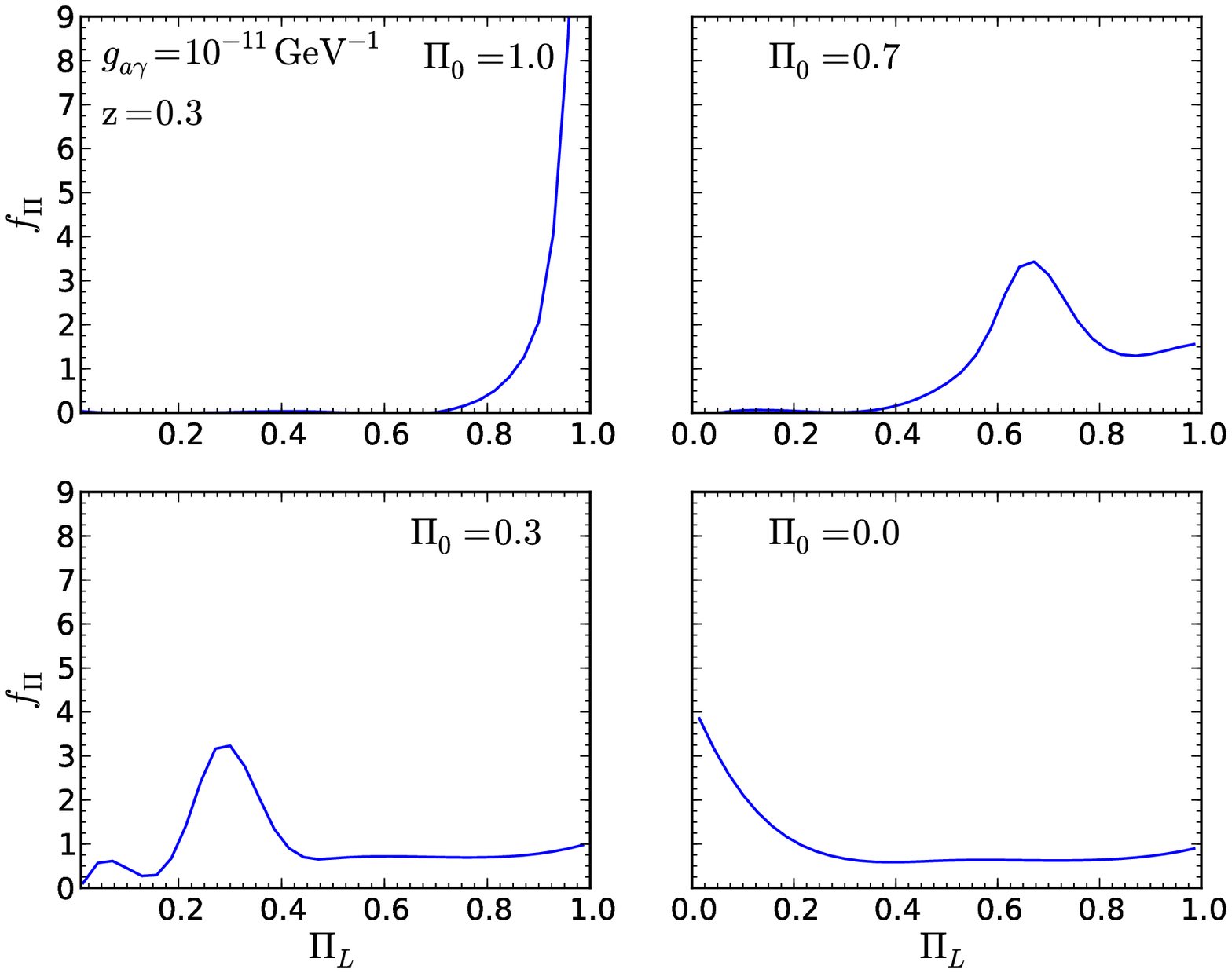, width =1.\columnwidth,  angle=0}
 \caption{
Probability density function $f_\Pi$ for the final linear polarization ${\Pi}_L$ after propagation in the extragalactic magnetic field, considering $10^4$ GRBs at redshift $z=0.3$, with initial linear polarization $\Pi_0=0.0,0.3,0.7,1.0$.
\label{fig4}}
\end{figure}
%%%%%%%%%%%%%%%%%%%%%%%%%%%%%%%%%%%%%%%%%%%%%%%%%%%%%%%%%%%%%%%%%%%%% 

%%%%%%%%%%%%%%%%%%%%%%%%%%% FIGURE 3 %%%%%%%%%%%%%%%%%%%%%%%%%%%%%%%%%
\begin{figure}[t]
\centering
\epsfig{figure=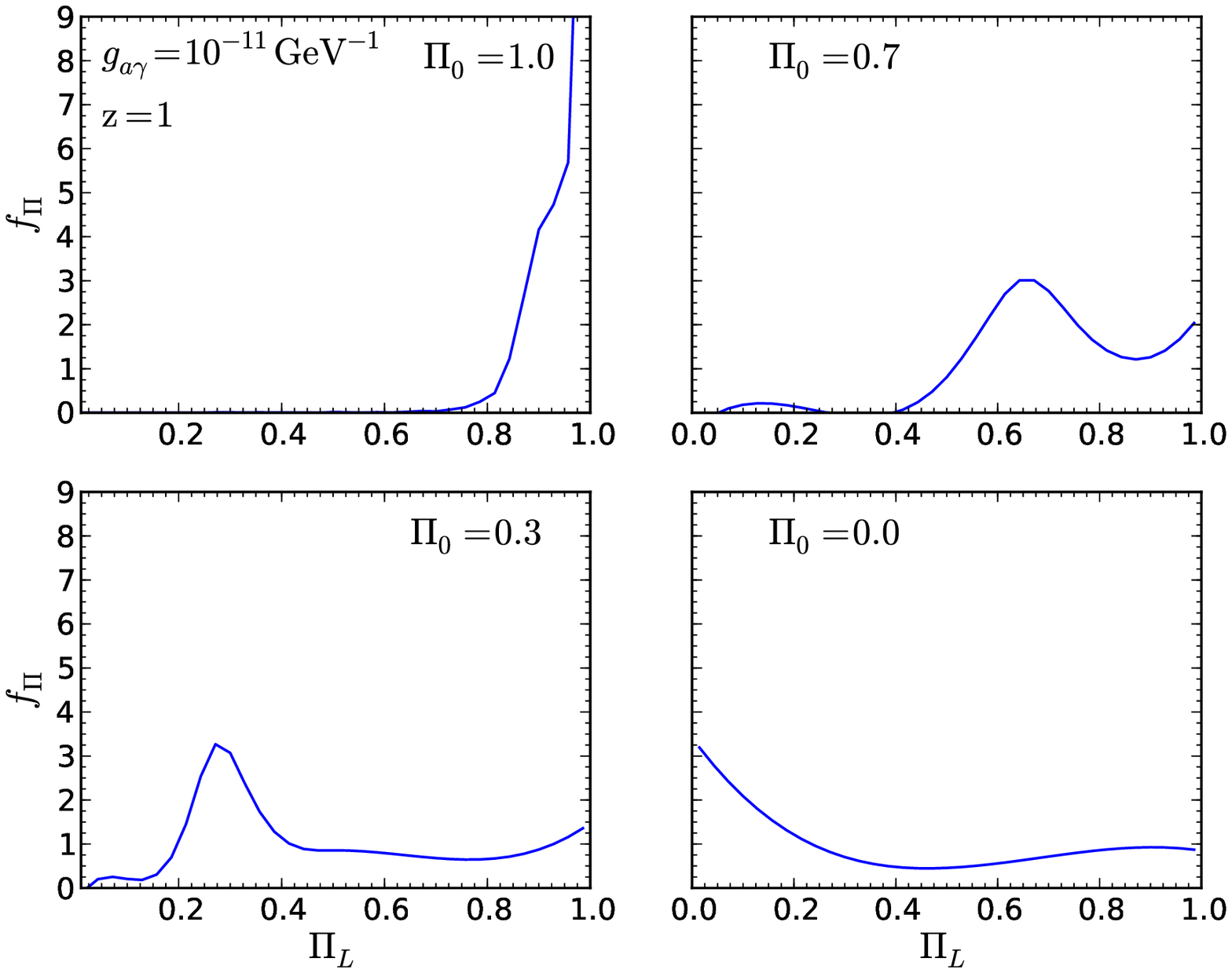, width =1.\columnwidth,  angle=0}
 \caption{
Probability density function $f_\Pi$ for the final linear polarization ${\Pi}_L$ after propagation in the extragalactic magnetic field, considering $10^4$ GRBs at redshift $z=1$, with initial linear polarization $\Pi_0=0.0,0.3,0.7,1.0$.
\label{fig5}}
\end{figure}
%%%%%%%%%%%%%%%%%%%%%%%%%%%%%%%%%%%%%%%%%%%%%%%%%%%%%%%%%%%%%%%%%%%%% 

%%%%%%%%%%%%%%%%%%%%%%%%%%% FIGURE 3 %%%%%%%%%%%%%%%%%%%%%%%%%%%%%%%%%
\begin{figure}[t]
\centering
\epsfig{figure=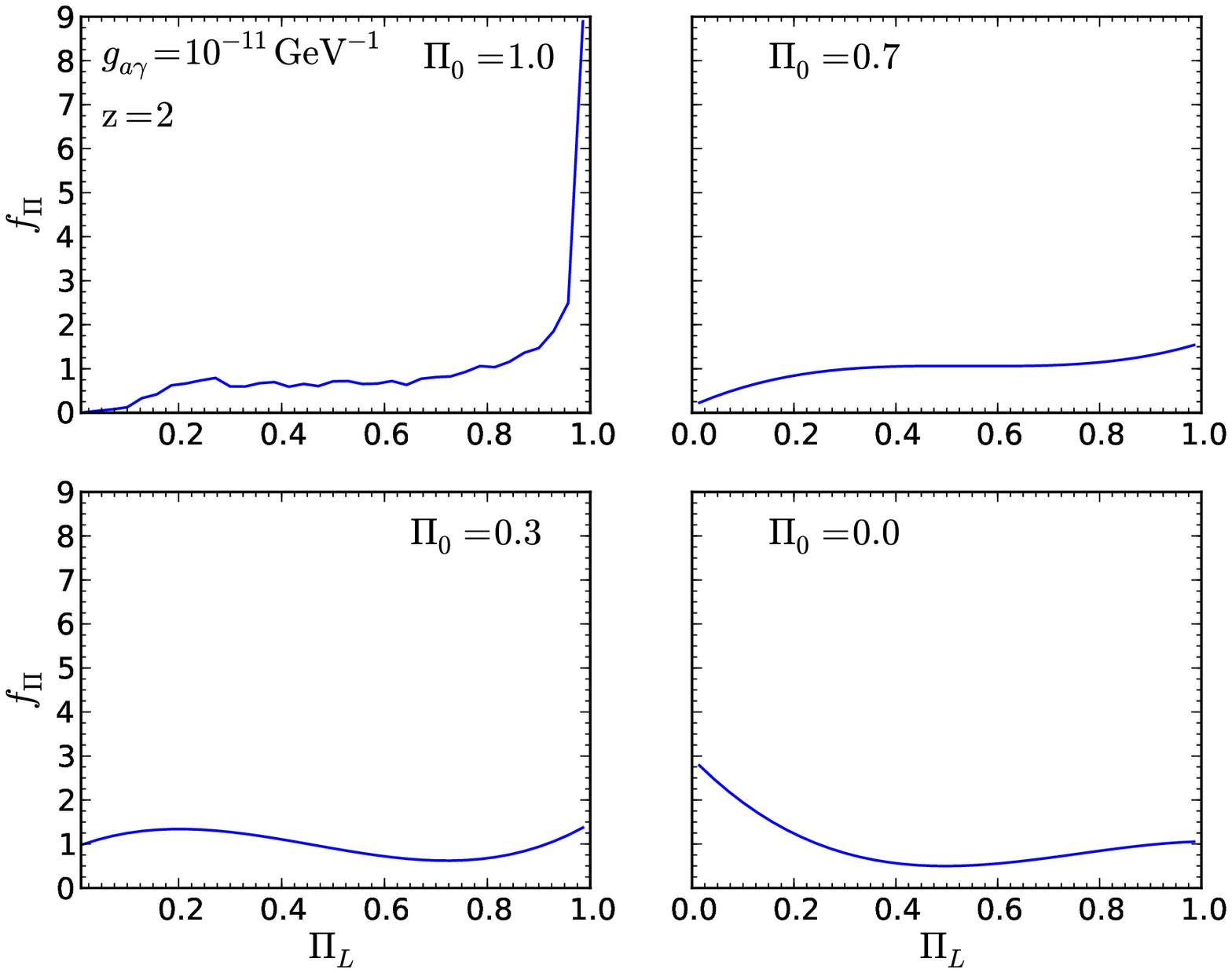, width =1.\columnwidth,  angle=0}
 \caption{
Probability density function $f_\Pi$ for the final linear polarization ${\Pi}_L$ after propagation in the extragalactic magnetic field, considering $10^4$ GRBs at redshift $z=2$, with initial linear polarization $\Pi_0=0.0,0.3,0.7,1.0$.
\label{fig6}}
\end{figure}
%%%%%%%%%%%%%%%%%%%%%%%%%%%%%%%%%%%%%%%%%%%%%%%%%%%%%%%%%%%%%%%%%%%%% 

%----------------------------------------------
\subsection{Intracluster magnetic fields} 
%----------------------------------------------
Given that GRBs are at cosmological distances, there is a nontrivial chance that in some cases their line of sight crosses a cluster of galaxies. So, it is worthwhile to investigate what happens in this instance. 

Observations have shown that the presence of magnetic fields with average strength $B \simeq 10^{- 6} \, {\rm G}$ is a typical feature of the 
intracluster region. Even more remarkable is the fact that observations are able to yield information about the associated coherence length, which turns out to be $L \simeq 10$~kpc~\cite{carilli}. As in the case of extragalactic magnetic fields, we assume a cellular structure for the intracluster fields, with domain size \mbox{$L \simeq 10$~kpc}. As before, plasma effects are expected to be present. Specifically, the electron density in the intracluster medium is \mbox{$n_e \simeq 1.0 \times 10^{-3} \, {\rm cm}^{-3}$}~\cite{Dolag:2004kp}, yielding a plasma frequency \mbox{${\omega}_{\rm pl} \simeq 1.2 \times 10^{-12} \, {\rm eV}$}. 

With these input values, the low-energy cut-off $E_L$ in Eq.~(\ref{eq:EL}) is $E_L \simeq 3.6 \, {\rm keV}$ for $m < 10^{- 12} \, {\rm eV}$, thereby guaranteeing that for our benchmark values $E=100$~keV and $m = 10^{- 13} \, {\rm eV}$  we are in the strong-mixing regime, where the photon-ALP conversion probability is energy-independent. 

In Figure~7 we plot the average linear polarization ${\Pi}_L$ and the photon survival probability $P_{\gamma \gamma}$ as a function of the distance $D$ traveled inside the intracluster region, assuming that the source is fully polarized along the $x$ direction. The averaging process is again over an ensemble of $10^4$ realizations of the random magnetic field network. We find that for $D\gtrsim 200$~kpc both the polarization and the survival probability saturate at their limiting values,  $P_{\gamma \gamma} =2/3$ and  $\Pi_L \simeq 0.79$ respectively.

As before, the source dimming turns out to be unimportant. Thus, the result for the average final polarization shown in Figure~7 implies that whenever the line of sight to a GRB crosses a cluster of galaxies the polarization distributions shown in Figures~3,~4,~5 and 6 are expected to be further smeared out.

%%%%%%%%%%%%%%%%%%%%%%%%%%% FIGURE 5 %%%%%%%%%%%%%%%%%%%%%%%%%%%%%%%%%
\begin{figure}[t]
\centering
\epsfig{figure=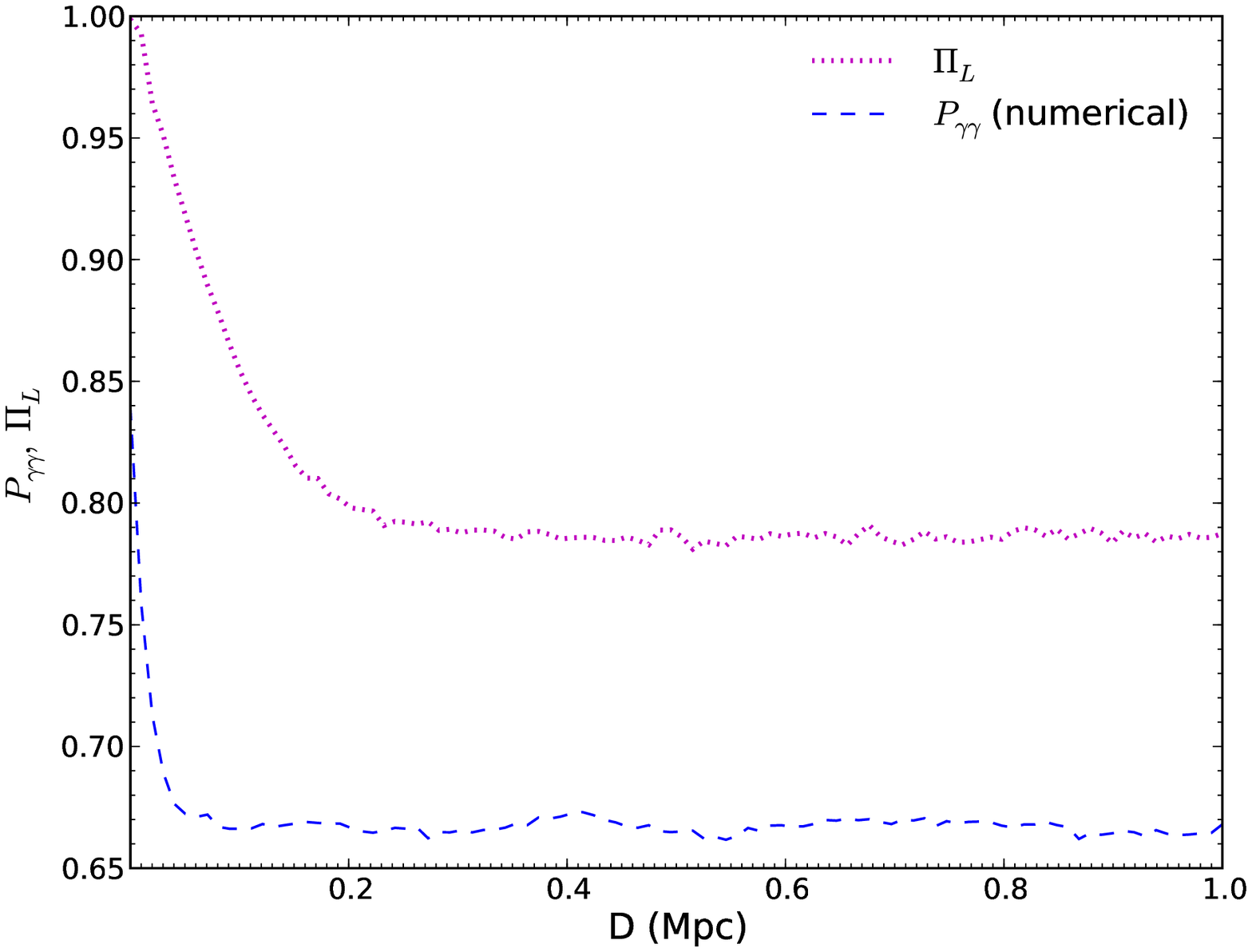, width =1.\columnwidth,  angle=0}
 \caption{
Average  linear polarization ${\Pi}_L$ (dotted line) and photon survival probability $P_{\gamma \gamma}$ 
(dashed line)  as a function of the distance $D$ traveled inside the intracluster region. 
The source is assumed to be fully polarized along the $x$ direction.

\label{fig7}}
\end{figure}
%%%%%%%%%%%%%%%%%%%%%%%%%%%%%%%%%%%%%%%%%%%%%%%%%%%%%%%%%%%%%%%%%%%%% 

%%%%%%%%%%%%%%%%%%%%%%%%%%% FIGURE 5 %%%%%%%%%%%%%%%%%%%%%%%%%%%%%%%%%
\begin{figure}[t]
\centering
\epsfig{figure=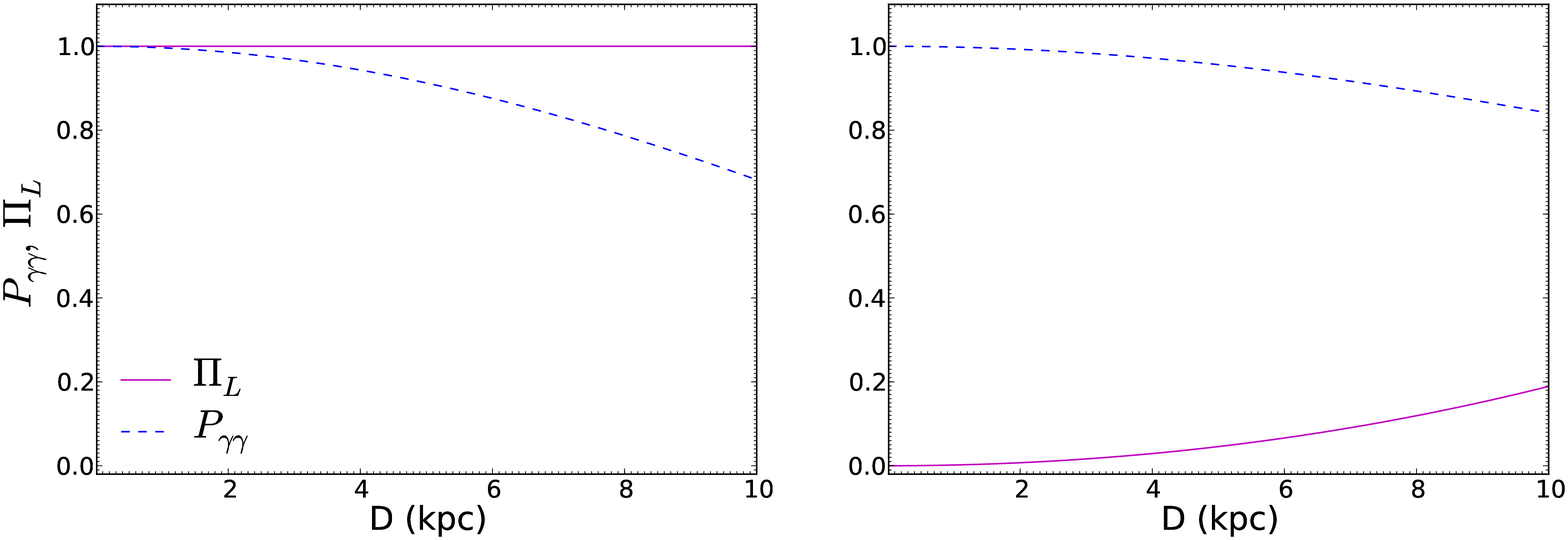, width =1.\columnwidth,  angle=0}
 \caption{
Linear polarization ${\Pi}_L$  (solid curve) and survival probability $P_{\gamma \gamma}$ (dashed curve) for the photon-ALP mixing in the regular component of the Galactic magnetic field as a function of the photon path $D$ in the Galaxy. We have assumed photons to be fully polarized along the $y$ direction (left panel) or completely unpolarized (right panel) before entering the Galaxy. We have taken as Galactic magnetic field $B= 4 \times 10^{-6}$~G, oriented
along the $y$ axis.
\label{fig8}}
\end{figure}
%%%%%%%%%%%%%%%%%%%%%%%%%%%%%%%%%%%%%%%%%%%%%%%%%%%%%%%%%%%%%%%%%%%%% 

%----------------------------------------------
\subsection{Galactic magnetic fields} 
%----------------------------------------------

Finally, the beam from a GRB obviously crosses the Milky Way before being detected. Observations over the last three decades have led to a rather detailed picture of the Milky Way magnetic field. Perhaps, its most important feature is that it consists of two components, a regular and a turbulent one. We proceed to address the ALP-induced polarization and the photon-ALP conversion in these two magnetic field configurations.

{\it Regular component} -- Measurements of the Faraday rotation based on pulsar observations have shown that this component is parallel to the Galactic plane. Its strength varies between $B \simeq 2 \times 10^{- 6} \, {\rm G}$ in the Solar neighborhood and $B \simeq 4 \times 10^{- 6} \, {\rm G}$ at 3 kpc from the center~\cite{han}. Moreover, the associated coherence length is $L \simeq10 \, {\rm kpc}$. Inside the Milky Way disk the electron density is \mbox{$n_e \simeq 1.1 \times 10^{-2} \, {\rm cm}^{-3}$}~\cite{Digel}, resulting in a plasma frequency \mbox{${\omega}_{\rm pl} \times 4.1 \times 10^{-12} \, {\rm eV}$}. Assuming the whole Galaxy as a single domain with an homogeneous magnetic field, for these input values Eq.~(\ref{eq:EL}) entails $E_L \simeq 10 \, {\rm keV}$ for 
$m < 10^{- 12} \, {\rm eV}$, so that for our reference values $E = 100 \, {\rm keV}$ and $m = 10^{- 13} \, {\rm eV}$ we are in the strong-mixing regime. In 
Figure~8 we show the linear polarization ${\Pi}_L$ and the photon survival probability $P_{\gamma \gamma}$ as a function of the photon path $D$ in the Galaxy, taking as Galactic magnetic field $B \simeq 4 \times 10^{- 6} \, {\rm G}$, oriented along the $y$ direction. Both quantities depend on the initial photon polarization. In the left panel we have chosen the initial linear polarization along the $y$ axis. In this situation, we find a final survival probability $P_{\gamma \gamma} \simeq 0.68$, while the initial linear polarization $\Pi_L=1$ remains unchanged. Instead, in the right panel we have started our simulation with a completely unpolarized state ($\Pi_L=0$). In this case, the photon-ALP conversion produces seizable effects on both the photon flux and the photon polarization. In particular, the  survival probability after Galaxy crossing is $P_{\gamma \gamma} \simeq 0.84$ and the final linear polarization $\Pi_L \simeq 0.19$. A detailed study of this effect would require a better modelling of the Galactic magnetic field, as in the study of photon-ALP conversion in the Milky Way performed in~\cite{Simet:2007sa}. However, we leave further refinements of these estimates for future work. 

{\it Turbulent component} -- Over much smaller scales, the dominant Galactic magnetic field appears to be stochastic, with a Kolmogorov spectrum $\alpha = 5/3$~\cite{ferriere}. In practice, this component can be described by a cellular structure, with strength $B \simeq 1 \times 10^{- 6} \, {\rm G}$ and domain size 
$L \simeq 10^{- 2} \, {\rm pc}$. It is straightforward to realize that in this case the oscillation length $l_{\rm osc}=2\pi/\Delta_{\rm osc}$ is much larger 
than the domain size $L$, so that the photon-ALP conversion is vanishingly small.

\section{Conclusions}

Measuring the polarization of prompt GRB emission in the keV-MeV range represents one of the main challenges for high-energy astronomy 
of the next decade. Various polarimetric missions are currently being developed, which are expected to collect an all-sky rich sample of GRBs so as to allow for a meaningful statistical analysis of their polarization properties. As recently realized, important conclusions concerning the GRB emission models are expected to be drawn from these studies.

We have shown that the existence of ALPs with parameters lying in experimentally allowed ranges drastically modifies the GRB polarization pattern.
More specifically, cosmic magnetic fields of extragalactic, intracluster and Galactic origin along the line of sight to a GRB act as catalysts for significant photon-ALP mixing. In particular, due to the random structure of the extragalactic and intracluster magnetic fields, the amount of photon-ALP mixing strongly depends  on the orientation of the line of sight. Therefore, starting with a given source polarization a broad statistical distribution is expected to be detected when observing GRBs from different directions in the sky. This effect should be superimposed on the original distribution of the GRB polarization as predicted by standard emission models.  As a result, the scatter in the distribution arising from the photon-ALP mixing hinders the possibility to extract information on the initial polarization for the observed GRBs. Alternatively, in the lack of any standard explanation, the observation of these peculiar broad distributions would hint at the existence of very light ALPs. We have restricted throughout our attention to the linear polarization because in all cases that we have analyzed the circular polarization turns out to be vanishingly small. This circumstance is at variance with the case of chameleontic ALPs~\cite{Burrage:2008ii}, and so it can be used to further investigate 
the nature of ALPs.

In the last few years, different hints and possible signatures  of very light ALPs have been proposed in connection with astrophysical observations~\cite{De Angelis:2007yu,Bassan:2009gy,MirizziHigh,Simet:2007sa,Hochmuth:2007hk,De Angelis:2007dy,De Angelis:2007dya,Hooper:2007bq,Burrage:2009mj,Hutsemekers:2008iv,Chelouche:2009iv}. In general, the presence of ALPs can play a role for values of the photon-ALP coupling constant $g_{a\gamma} \lesssim 10^{-11}$~GeV$^{-1}$, namely one order of magnitude lower than the current experimental limit set by the CAST experiment. This indicates the high potential of astrophysical observations to probe ALPs in a region of the their parameter space otherwise unreachable. It remains to see whether the above hints at ALPs will ultimately converge in a robust signature. Every astrophysical argument has its own systematic uncertainties and its own recognized or unrecognized loop-holes, so that in order to corner the ALPs it is important to use as many independent channels and as many different approaches as possible. In this sense, we believe that the present study of ALP effects on the GRB polarization should be regarded as an additional opportunity to uncover the existence of very light ALPs through astrophysical observations. Remarkably enough, no dedicated experiment is required and our predictions can be easily tested in the near future with the planned polarimetric missions for GRBs.

%%%%%%%%%%%%%%%%%%%%%%%%%%%%%%%%%%%%%%%%%%%%%%%%%%%%%%%%%%%%%%%%%%%%%%
\section*{Acknowledgments} %%%%%%%%%%%%%%%%%%%%%%%%%%%%%%%%%%%%%%%%%%%%%%%%
%%%%%%%%%%%%%%%%%%%%%%%%%%%%%%%%%%%%%%%%%%%%%%%%%%%%%%%%%%%%%%%%%%%%%%
We would like to thank Enrico Costa, Gabriele Ghisellini and Kenji Toma for discussions about the possibility to measure the polarization of distant high-energy astrophysical sources in a systematic fashion. M. R. also thanks the Dipartimento di Fisica Nucleare e Teorica, Universit\`a di Pavia, for support. 

%%%%%%%%%%%%%%%%%%%%%%%%%%%%%%%%%%%%%%%%%%%%%%%%%%%%%%%%%%%%%%%%%%%%%%
\section*{References} %%%%%%%%%%%%%%%%%%%%%%%%%%%%%%%%%%%%%%%%%%%%%%%%
%%%%%%%%%%%%%%%%%%%%%%%%%%%%%%%%%%%%%%%%%%%%%%%%%%%%%%%%%%%%%%%%%%%%%%


\begin{thebibliography}{99}

  



\bibitem{masso1}
  E.~Masso and R.~Toldra,
  ``On a Light Spinless Particle Coupled to Photons,''
  Phys.\ Rev.\  D {\bf 52}, 1755 (1995) 
  [hep-ph/9503293].
  %%CITATION = PHRVA,D52,1755;%%




\bibitem{masso2}
E. Masso and R. Toldra, 
  E.~Masso and R.~Toldra,
  ``New constraints on a light spinless particle coupled to photons,''
  Phys.\ Rev.\  D {\bf 55}, 7967 (1997)
  [hep-ph/9702275].
  %%CITATION = PHRVA,D55,7967;%%


\bibitem{coriano1}
  C.~Coriano and N.~Irges,
  ``Windows over a new low energy axion,''
  Phys.\ Lett.\  B {\bf 651}, 298 (2007)
  [hep-ph/0612140].
  %%CITATION = PHLTA,B651,298;%%







\bibitem{coriano2}
C.~Coriano, N.~Irges and S.~Morelli, 
  C.~Coriano, N.~Irges and S.~Morelli,
  ``Stueckelberg axions and the effective action of anomalous Abelian models.
  I: A unitarity analysis of the Higgs-axion mixing,''
  JHEP {\bf 0707}, 008 (2007)
  [hep-ph/0701010].
  %%CITATION = JHEPA,0707,008;%%




\bibitem{kk}
 S.~Chang, S.~Tazawa and M.~Yamaguchi,
  ``Axion model in extra dimensions with TeV scale gravity,''
  Phys.\ Rev.\  D {\bf 61}, 084005 (2000)
  [hep-ph/9908515].
  %%CITATION = PHRVA,D61,084005;%%


%\cite{Svrcek:2006yi}
\bibitem{Svrcek:2006yi}
  P.~Svrcek and E.~Witten,
  ``Axions in string theory,''
  JHEP {\bf 0606}, 051 (2006)
  [hep-th/0605206].
  %%CITATION = JHEPA,0606,051;%%

%\cite{Arvanitaki:2009fg}
\bibitem{Arvanitaki:2009fg}
  A.~Arvanitaki, S.~Dimopoulos, S.~Dubovsky, N.~Kaloper and J.~March-Russell,
  ``String Axiverse,''
  arXiv:0905.4720 [hep-th].
  %%CITATION = ARXIV:0905.4720;%%

%\cite{Peccei:1977hh}
\bibitem{Peccei:1977hh}
  R.~D.~Peccei and H.~R.~Quinn,
  ``CP Conservation In The Presence Of Instantons,''
  Phys.\ Rev.\ Lett.\  {\bf 38}, 1440 (1977).
  %%CITATION = PRLTA,38,1440;%%


%\cite{Peccei:1977ur}
\bibitem{Peccei:1977ur}
  R.~D.~Peccei and H.~R.~Quinn,
  ``Constraints Imposed By CP Conservation In The Presence Of Instantons,''
  Phys.\ Rev.\  D {\bf 16}, 1791 (1977).
  %%CITATION = PHRVA,D16,1791;%%

%\cite{Weinberg:1977ma}
\bibitem{Weinberg:1977ma}
  S.~Weinberg,
  ``A New Light Boson?,''
  Phys.\ Rev.\ Lett.\  {\bf 40}, 223 (1978).
  %%CITATION = PRLTA,40,223;%%

%\cite{Wilczek:1977pj}
\bibitem{Wilczek:1977pj}
  F.~Wilczek,
  ``Problem Of Strong P And T Invariance In The Presence Of Instantons,''
  Phys.\ Rev.\ Lett.\  {\bf 40}, 279 (1978).
  %%CITATION = PRLTA,40,279;%%

%\cite{Kim:2008hd}
\bibitem{Kim:2008hd}
  J.~E.~Kim and G.~Carosi,
  ``Axions and the Strong CP Problem,''
  arXiv:0807.3125 [hep-ph].
  %%CITATION = ARXIV:0807.3125;%%

%\cite{Masso:2006id}
\bibitem{Masso:2006id}
  E.~Masso,
  ``Axions and their relatives,''
  Lect.\ Notes Phys.\  {\bf 741}, 83 (2008) 
  [hep-ph/0607215].
  %%CITATION = LNPHA,741,83;%%

\bibitem{cdm}
  P.~Sikivie,
  ``Axion cosmology,''
  Lect.\ Notes Phys.\  {\bf 741}, 19 (2008)
  [astro-ph/0610440].
  %%CITATION = LNPHA,741,19;%%



\bibitem{carroll}
  S.~M.~Carroll,
  ``Quintessence and the rest of the world,''
  Phys.\ Rev.\ Lett.\  {\bf 81}, 3067 (1998)
  [astro-ph/9806099].
  %%CITATION = PRLTA,81,3067;%%

%\cite{Jaeckel:2010ni}
\bibitem{Jaeckel:2010ni}
  J.~Jaeckel and A.~Ringwald,
  ``The Low-Energy Frontier of Particle Physics,''
  arXiv:1002.0329 [hep-ph].
  %%CITATION = ARXIV:1002.0329;%%

\bibitem{Raffelt:1987im}
  G.~Raffelt and L.~Stodolsky,
  ``Mixing of the photon with low mass particles,''
  Phys.\ Rev.\ D {\bf 37}, 1237 (1988).
  %%CITATION = PHRVA,D37,1237;%%

\bibitem{sikivie}
  P.~Sikivie,
  ``Experimental tests of the ``invisible'' axion,''
  Phys.\ Rev.\ Lett.\  {\bf 51}, 1415 (1983), 
  Erratum {\it ibid.} {\bf 52}, 695 (1984).
  %%CITATION = PRLTA,51,1415;%%

\bibitem{Anselm:1987vj}
  A.~A.~Anselm,
  ``Experimental test for arion $\leftrightarrow$ photon 
  oscillations in a homogeneous constant magnetic field,''
  Phys.\ Rev.\ D {\bf 37}, 2001 (1988).





%\cite{Duffy:2006aa}
\bibitem{Duffy:2006aa}
  L.~D.~Duffy {\it et al.},
  ``A High Resolution Search for Dark-Matter Axions,''
  Phys.\ Rev.\  D {\bf 74}, 012006 (2006)
  [astro-ph/0603108].
  %%CITATION = PHRVA,D74,012006;%%

%\cite{Zioutas:2004hi}
\bibitem{Zioutas:2004hi}
  K.~Zioutas {\it et al.}  [CAST Collaboration],
  ``First results from the CERN Axion Solar Telescope (CAST),''
  Phys.\ Rev.\ Lett.\  {\bf 94}, 121301 (2005)
  [hep-ex/0411033].
  %%CITATION = PRLTA,94,121301;%%

%\cite{Andriamonje:2007ew}
\bibitem{Andriamonje:2007ew}
  S.~Andriamonje {\it et al.}  [CAST Collaboration],
  ``An improved limit on the axion-photon coupling from the CAST experiment,''
  JCAP {\bf 0704}, 010 (2007)
  [hep-ex/0702006].
  %%CITATION = JCAPA,0704,010;%%

%\cite{Arik:2008mq}
\bibitem{Arik:2008mq}
  E.~Arik {\it et al.}  [CAST Collaboration],
  ``Probing eV-scale axions with CAST,''
  JCAP {\bf 0902}, 008 (2009)
  [arXiv:0810.4482 [hep-ex]].
  %%CITATION = JCAPA,0902,008;%%


%\cite{Robilliard:2007bq}
\bibitem{Robilliard:2007bq}
  C.~Robilliard, R.~Battesti, M.~Fouche, J.~Mauchain, A.~M.~Sautivet, F.~Amiranoff and C.~Rizzo,
  ``No light shining through a wall,''
  Phys.\ Rev.\ Lett.\  {\bf 99}, 190403 (2007)
  [arXiv:0707.1296 [hep-ex]].
  %%CITATION = PRLTA,99,190403;%%

%\cite{Chou:2007zzc}
\bibitem{Chou:2007zzc}
  A.~S.~Chou {\it et al.}  [GammeV (T-969) Collaboration],
  ``Search for axion-like particles using a variable baseline photon
  regeneration technique,''
  Phys.\ Rev.\ Lett.\  {\bf 100}, 080402 (2008)
  [arXiv:0710.3783 [hep-ex]].
  %%CITATION = PRLTA,100,080402;%%


%\cite{Afanasev:2008jt}
\bibitem{Afanasev:2008jt}
  A.~Afanasev {\it et al.},
  ``New Experimental limit on Optical Photon Coupling to Neutral, Scalar
  Bosons,''
  Phys.\ Rev.\ Lett.\  {\bf 101}, 120401 (2008)
  [arXiv:0806.2631 [hep-ex]].
  %%CITATION = PRLTA,101,120401;%%

%\cite{Fouche:2008jk}
\bibitem{Fouche:2008jk}
  M.~Fouche {\it et al.},
  ``Search for photon oscillations into massive particles,''
  Phys.\ Rev.\  D {\bf 78}, 032013 (2008)
  [arXiv:0808.2800 [hep-ex]].
  %%CITATION = PHRVA,D78,032013;%%

%\cite{Ehret:2009sq}
\bibitem{Ehret:2009sq}
  K.~Ehret {\it et al.}  [ALPS collaboration],
  ``Resonant laser power build-up in ALPS - a 'light-shining-through-walls'
  experiment -,''
  Nucl.\ Instrum.\ Meth.\  A {\bf 612}, 83 (2009)
  [arXiv:0905.4159 [physics.ins-det]].
  %%CITATION = NUIMA,A612,83;%%

%\cite{Ehret:2010mh}
\bibitem{Ehret:2010mh}
  K.~Ehret {\it et al.},
  ``New ALPS Results on Hidden-Sector Lightweights,''
  arXiv:1004.1313 [hep-ex].
  %%CITATION = ARXIV:1004.1313;%%

  
  \bibitem{axion2010}
   A.~Ringwald,
  ``Challenges and Opportunities for the Next Generation of Photon Regeneration
  Experiments,''
  arXiv:1003.2339 [hep-ph].
  %%CITATION = ARXIV:1003.2339;%%



\bibitem{Maiani:1986md}
  L.~Maiani, R.~Petronzio and E.~Zavattini,
  ``Effects of nearly massless, spin zero particles
   on light propagation in a magnetic field,''
  Phys.\ Lett.\  B {\bf 175}, 359 (1986).
  %%CITATION = PHLTA,B175,359;%%

\bibitem{Gasperini:1987da}
  M.~Gasperini,
  ``Axion production by electromagnetic fields,''
  Phys.\ Rev.\ Lett.\  {\bf 59}, 396 (1987).
  %%CITATION = PRLTA,59,396;%%


%\cite{Zavattini:2005tm}
\bibitem{Zavattini:2005tm}
  E.~Zavattini {\it et al.}  [PVLAS Collaboration],
  ``Experimental observation of optical rotation generated in vacuum by a
  magnetic field,''
  Phys.\ Rev.\ Lett.\  {\bf 96}, 110406 (2006)
  [Erratum-ibid.\  {\bf 99}, 129901 (2007)]
  [hep-ex/0507107].
  %%CITATION = PRLTA,96,110406;%%


%\cite{Zavattini:2007ee}
\bibitem{Zavattini:2007ee}
  E.~Zavattini {\it et al.}  [PVLAS Collaboration],
  ``New PVLAS results and limits on magnetically induced optical rotation and
  ellipticity in vacuum,''
  Phys.\ Rev.\  D {\bf 77}, 032006 (2008)
  [arXiv:0706.3419 [hep-ex]].
  %%CITATION = PHRVA,D77,032006;%%
  
  %\cite{Burrage:2008ii}
\bibitem{Burrage:2008ii}
  C.~Burrage, A.~C.~Davis and D.~J.~Shaw,
  ``Detecting Chameleons: The Astronomical Polarization Produced by
  Chameleon-like Scalar Fields,''
  Phys.\ Rev.\  D {\bf 79}, 044028 (2009)
  [arXiv:0809.1763 [astro-ph]].
  %%CITATION = PHRVA,D79,044028;%%


\bibitem{Toma:2008vi}
  K.~Toma {\it et al.},
  ``Statistical Properties of Gamma-Ray Burst Polarization,''
  Astrophys.\ J.\ {\bf 698}, 1042 (2009)   
  [arXiv:0812.2483 [astro-ph]].
  %%CITATION = ARXIV:0812.2483;%%
  
  \bibitem{harari}
  D.~Harari and P.~Sikivie,
  ``Effects of a Nambu-Goldstone boson on the polarization of radio galaxies
  and the cosmic microwave background,''
  Phys.\ Lett.\  B {\bf 289}, 67 (1992).
  %%CITATION = PHLTA,B289,67;%%

  
  
  \bibitem{poet}
    J.~E.~Hill {\it et al.},
  ``POET: POlarimeters for Energetic Transients,''
  AIP Conf.\ Proc.\  {\bf 1065}, 331 (2008)
  [arXiv:0810.2499 [astro-ph]].
  %%CITATION = APCPC,1065,331;%%


    
  %\cite{Mizuno:2004ag}
\bibitem{Mizuno:2004ag}
  T.~Mizuno {\it et al.},
  ``Beam Test of a Prototype Detector Array for the PoGO Astronomical Hard
  X-Ray/Soft Gamma-Ray Polarimeter,''
  Nucl.\ Instrum.\ Meth.\  A {\bf 540}, 158 (2005)
  [astro-ph/0411341].
  %%CITATION = NUIMA,A540,158;%%
  
  %\cite{Produit:2005zu}
\bibitem{Produit:2005zu}
  N.~Produit {\it et al.},
  ``POLAR, a compact detector for Gamma Ray Bursts photon polarization
  measurements,''
  Nucl.\ Instrum.\ Meth.\  A {\bf 550}, 616 (2005)
  [arXiv:astro-ph/0504605].
  %%CITATION = NUIMA,A550,616;%%

%\cite{Jahoda:2007pd}
\bibitem{Jahoda:2007pd}
  K.~Jahoda, K.~Black, P.~Deines-Jones, J.~E.~Hill, T.~Kallman, T.~Strohmayer and J.~H.~Swank,
  ``An X-ray Polarimeter for Constellation-X,''
  astro-ph/0701090.
  %%CITATION = ASTRO-PH/0701090;%%



%\cite{Costa:2008pb}
\bibitem{Costa:2008pb}
  E.~Costa {\it et al.},
  ``XPOL: a photoelectric polarimeter onboard XEUS,''
  arXiv:0810.2700 [astro-ph].
  %%CITATION = ARXIV:0810.2700;%%

%\cite{Greiner:2008yd}
\bibitem{Greiner:2008yd}
  J.~Greiner,
  ``GRIPS - Gamma-Ray Burst Investigation via Polarimetry and Spectroscopy,''
  Exper.\ Astron.\  {\bf 23}, 91 (2009)
  [arXiv:0808.0267 [astro-ph]].
  %%CITATION = EXASE,23,91;%%

%\cite{Tagliaferri:2010wk}
\bibitem{Tagliaferri:2010wk}
  G.~Tagliaferri {\it et al.},
  ``The New Hard X-ray Mission,''
  arXiv:1004.2691 [astro-ph.IM].
  %%CITATION = ARXIV:1004.2691;%%


%\cite{Mirizzi:2006zy}
\bibitem{Mirizzi:2006zy}
  A.~Mirizzi, G.~G.~Raffelt and P.~D.~Serpico,
  ``Photon axion conversion in intergalactic magnetic fields and  cosmological
  consequences,''
  Lect.\ Notes Phys.\  {\bf 741}, 115 (2008)
  [astro-ph/0607415].
  %%CITATION = LNPHA,741,115;%%

%\cite{Kronberg:1993vk}
\bibitem{Kronberg:1993vk}
  P.~P.~Kronberg,
  ``Extragalactic magnetic fields,''
  Rept.\ Prog.\ Phys.\  {\bf 57}, 325 (1994).
  %%CITATION = RPPHA,57,325;%%

%\cite{Grasso:2000wj}
\bibitem{Grasso:2000wj}
D.~Grasso and H.~R.~Rubinstein,
``Magnetic fields in the early universe,''
Phys.\ Rept.\  {\bf 348}, 163 (2001)
[astro-ph/0009061].
%%CITATION = PRPLC,348,163;%%


%\cite{Hinshaw:2008kr}
\bibitem{Hinshaw:2008kr}
WMAP Collaboration, G.~Hinshaw {\it et al.},
``Five-Year Wilkinson Microwave Anisotropy Probe
Observations:Data Processing, Sky Maps, \& Basic Results,''
Astrophys.\ J.\ Suppl.\  {\bf 180}, 225 (2009)
[arXiv:0803.0732 [astro-ph]].


\bibitem{Blasi:1999hu}
P.~Blasi, S.~Burles and A.~V.~Olinto,
``Cosmological Magnetic Fields Limits in an Inhomogeneous Universe,''
Astrophys.\ J.\  {\bf 514}, L79 (1999)
[astro-ph/9812487].
%%CITATION = ASJOA,514,L79;%%


%\cite{Raffelt:2006cw}
\bibitem{Raffelt:2006cw}
  G.~G.~Raffelt,
  ``Astrophysical axion bounds,''
  Lect.\ Notes Phys.\  {\bf 741}, 51 (2008)
  [hep-ph/0611350].
  %%CITATION = LNPHA,741,51;%%

\bibitem{Brockway:1996yr}
  J.~W.~Brockway, E.~D.~Carlson and G.~G.~Raffelt,
  ``SN 1987A gamma-ray limits on the conversion of pseudoscalars,''
  Phys.\ Lett.\ B {\bf 383}, 439 (1996)
  [astro-ph/ 9605197].

\bibitem{Grifols:1996id}
  J.~A.~Grifols, E.~Mass\'o and R.~Toldr\`a,
  ``Gamma rays from SN~1987A due to pseudoscalar conversion,''
  Phys.\ Rev.\ Lett.\ {\bf 77}, 2372 (1996)
  [astro-ph/9606028].
  %%CITATION = ASTRO-PH 9606028;%%





%\cite{Brax:2007ak}
\bibitem{Brax:2007ak}
  P.~Brax, C.~van de Bruck and A.~C.~Davis,
  ``Compatibility of the chameleon-field model with fifth-force experiments,
  cosmology, and PVLAS and CAST results,''
  Phys.\ Rev.\ Lett.\  {\bf 99}, 121103 (2007)
  [hep-ph/0703243].
  %%CITATION = PRLTA,99,121103;%%


  

%\cite{De Angelis:2007yu}
\bibitem{De Angelis:2007yu}
  A.~De Angelis, O.~Mansutti and M.~Roncadelli,
  ``Axion-Like Particles, Cosmic Magnetic Fields and Gamma-Ray Astrophysics,''
  Phys.\ Lett.\  B {\bf 659}, 847 (2008)
  [arXiv:0707.2695 [astro-ph]].
  %%CITATION = PHLTA,B659,847;%%

%\cite{Bassan:2009gy}
\bibitem{Bassan:2009gy}
  N.~Bassan and M.~Roncadelli,
  ``Photon-axion conversion in Active Galactic Nuclei?,''
  arXiv:0905.3752 [astro-ph.HE].
  %%CITATION = ARXIV:0905.3752;%%

%\cite{Christensson:2002ig}
\bibitem{Christensson:2002ig}
  M.~Christensson and M.~Fairbairn,
  ``Photon axion mixing in an inhomogeneous universe,''
  Phys.\ Lett.\  B {\bf 565}, 10 (2003)
  [astro-ph/0207525].
  %%CITATION = PHLTA,B565,10;%%


\bibitem{kosowski}
A.~Kosowsky
``Introduction to Microwave Background Polarization,''
Ann.\ Phys. \ (N.Y.) {\bf 246}, 49 (1996)
[astro-ph/9904102]

 \bibitem{rybicki}
G.~B.~Rybicki and A.~P.~Lightman,
{\it Radiative Processes in Astrophysics}  
(Wiley, Nwe York, 1979).


%\cite{Amsler:2008zzb}
\bibitem{Amsler:2008zzb}
  C.~Amsler {\it et al.}  [Particle Data Group],
  ``Review of particle physics,''
  Phys.\ Lett.\  B {\bf 667}, 1 (2008).
  %%CITATION = PHLTA,B667,1;%%

%\cite{Csaki:2001jk}
\bibitem{Csaki:2001jk}
  C.~Csaki, N.~Kaloper and J.~Terning,
  ``Effects of the intergalactic plasma on supernova dimming via  photon axion
  oscillations,''
  Phys.\ Lett.\  B {\bf 535}, 33 (2002)
  [hep-ph/0112212].
  %%CITATION = PHLTA,B535,33;%%

%\cite{Valageas:1999wa}
\bibitem{Valageas:1999wa}
  P.~Valageas, R.~Schaeffer and J.~Silk,
  ``The redshift evolution of Lyman-$\alpha$ absorbers,''
  Astron.\ Astrophys.\  {\bf 345}, 691 (1999)
  [astro-ph/9903388].
  %%CITATION = AAEJA,345,691;%%


%\cite{Dave:1998gm}
\bibitem{Dave:1998gm}
  R.~Dave, L.~Hernquist, N.~Katz and D.~H.~Weinberg,
  ``The Low Redshift Lyman Alpha Forest in Cold Dark Matter Cosmologies,''
  Astrophys.\ J.\  {\bf 511}, 521 (1999)
  [astro-ph/9807177].
  %%CITATION = ASJOA,511,521;%%

%\cite{Schaye:2001me}
\bibitem{Schaye:2001me}
  J.~Schaye,
  ``Model-independent insights into the nature of the Lyman-alpha forest and
  the distribution of matter in the universe,''
  Astrophys.\ J.\  {\bf 559}, 507 (2001)
  [astro-ph/0104272].
  %%CITATION = ASJOA,559,507;%%


%\cite{Rubbia:2007hf}
\bibitem{Rubbia:2007hf}
  A.~Rubbia and A.~S.~Sakharov,
  ``Constraining axion by polarized prompt emission from gamma ray bursts,''
  Astropart.\ Phys.\  {\bf 29}, 20 (2008)
  [arXiv:0708.2646 [hep-ph]].
  %%CITATION = APHYE,29,20;%%

\bibitem{MirizziHigh}
A.~Mirizzi and D.~Montanino, ``Stochastic conversion of TeV photons into axion-like particles in
 extra-galactic magnetic fields,'' JCAP {\bf 0912}, 004 (2009)
 [arXiv:0911.0015 [astro-ph.HE]].

\bibitem{carilli}
  C.~L.~Carilli and G.~B.~Taylor,
  ``Cluster Magnetic Fields,''
  Ann.\ Rev.\ Astron.\ Astrophys.\  {\bf 40}, 319 (2002)
  [astro-ph/0110655].
  %%CITATION = ARAAA,40,319;%%

%\cite{Dolag:2004kp}
\bibitem{Dolag:2004kp}
  K.~Dolag, D.~Grasso, V.~Springel and I.~Tkachev,
  ``Constrained simulations of the magnetic field in the local universe and
  the propagation of UHECRs,''
  JCAP {\bf 0501}, 009 (2005)
  [astro-ph/0410419].
  %%CITATION = JCAPA,0501,009;%%

%\cite{han}
\bibitem{han}
  J.~L.~Han, R.~N.~Manchester, A.~G.~Lyne, G.~J.~Qiao and W.~van Straten,
  ``Pulsar rotation measures and the large-scale structure of Galactic magnetic
  field,''
  Astrophys.\ J.\  {\bf 642}, 868 (2006)
  [astro-ph/0601357].
  %%CITATION = ASJOA,642,868;%%

\bibitem{Digel}
R.~Almy, D.~McCammon, S.~W.~Digel, L.~Bronfman, J.~May, ``Distance Limits 
on the Bright X-Ray Emission Toward the Galactic Center: Evidence for a 
Very Hot Interstellar Medium in the Galactic X-Ray Bulge,'' 
  Astrophys.\ J.\  {\bf 545}, 290 (2000).
  
  
  %\cite{Simet:2007sa}
\bibitem{Simet:2007sa}
  M.~Simet, D.~Hooper and P.~D.~Serpico,
  ``The Milky Way as a Kiloparsec-Scale Axionscope,''
  Phys.\ Rev.\  D {\bf 77}, 063001 (2008)
  [arXiv:0712.2825 [astro-ph]].
  %%CITATION = PHRVA,D77,063001;%%


%\cite{ferriere}
\bibitem{ferriere}
  J.~L.~Han, K.~Ferriere and R.~N.~Manchester,
  ``The spatial energy spectrum of magnetic fields in our Galaxy,''
  Astrophys.\ J.\  {\bf 610}, 820 (2004)
  [astro-ph/0404221].
  %%CITATION = ASJOA,610,820;%%

  
 %\cite{Hochmuth:2007hk}
\bibitem{Hochmuth:2007hk}
  K.~A.~Hochmuth and G.~Sigl,
  ``Effects of Axion-Photon Mixing on Gamma-Ray Spectra from Magnetized
  Astrophysical Sources,''
  Phys.\ Rev.\  D {\bf 76}, 123011 (2007)
  [arXiv:0708.1144 [astro-ph]].
  %%CITATION = PHRVA,D76,123011;%%

%\cite{De Angelis:2007dy}
\bibitem{De Angelis:2007dy}
  A.~De Angelis, O.~Mansutti and M.~Roncadelli,
  ``Evidence for a new light spin-zero boson from cosmological gamma-ray
  propagation?,''
  Phys.\ Rev.\  D {\bf 76}, 121301 (2007)
  [arXiv:0707.4312 [astro-ph]].
  %%CITATION = PHRVA,D76,121301;%%

\bibitem{De Angelis:2007dya}
  A.~De Angelis, O.~Mansutti, M.~Persic and M.~Roncadelli,
  ``Photon propagation and the very high energy $\gamma$-ray spectra of blazars: 
  how transparent is the Universe?,'' 
  Mon.\ Not.\ R.\ Astron. \ Soc. \ D {\bf 394}, L21 (2009)
  [arXiv:0807.4246 [astro-ph]].
  
%\cite{Hooper:2007bq}
\bibitem{Hooper:2007bq}
  D.~Hooper and P.~D.~Serpico,
  ``Detecting Axion-Like Particles With Gamma Ray Telescopes,''
  Phys.\ Rev.\ Lett.\  {\bf 99}, 231102 (2007)
  [arXiv:0706.3203 [hep-ph]].
  %%CITATION = PRLTA,99,231102;%%

  %\cite{Burrage:2009mj}
\bibitem{Burrage:2009mj}
  C.~Burrage, A.~C.~Davis and D.~J.~Shaw,
  ``Active Galactic Nuclei Shed Light on Axion-like-Particles,''
  Phys.\ Rev.\ Lett.\  {\bf 102}, 201101 (2009)
  [arXiv:0902.2320 [astro-ph.CO]].
  %%CITATION = PRLTA,102,201101;%%

%\cite{Hutsemekers:2008iv}
\bibitem{Hutsemekers:2008iv}
  D.~Hutsemekers, A.~Payez, R.~Cabanac, H.~Lamy, D.~Sluse, B.~Borguet and J.~R.~Cudell,
  ``Large-Scale Alignments of Quasar Polarization Vectors: Evidence at
  Cosmological Scales for Very Light Pseudoscalar Particles Mixing with
  Photons?,''
  arXiv:0809.3088 [astro-ph].
  %%CITATION = ARXIV:0809.3088;%%

%\cite{Chelouche:2009iv}
\bibitem{Chelouche:2009iv}
  D.~Chelouche, R.~Rabadan, S.~Pavlov,  and F.~Castejon,
  ``Spectral Signatures of Photon-Particle Oscillations from Celestial Objects,''
  Astrophys.\ J.\ Suppl.\ {\bf 180}, 1 (2009)
  [arXiv:0806.0411 [astro-ph]].
  %%CITATION = ARXIV:0809.3088;%%







\end{thebibliography}
\end{document}